\documentclass[pdflatex,sn-mathphys-num]{sn-jnl}



\usepackage{graphicx}%
\usepackage{multirow}%
\usepackage{amsmath,amssymb,amsfonts}%
\usepackage{amsthm}%
\usepackage{mathrsfs}%
\usepackage[title]{appendix}%
\usepackage{xcolor}%
\usepackage{textcomp}%
\usepackage{manyfoot}%
\usepackage{booktabs}%
\usepackage[utf8]{inputenc}
\usepackage{algorithm}%
\usepackage{algorithmicx}%
\usepackage{algpseudocode}%
\usepackage{listings}%
\usepackage[utf8]{inputenc}%
\usepackage{soul}
\usepackage{graphicx}
\usepackage[backend=biber,style=nature]{biblatex} 
\usepackage{float}
\usepackage{placeins}
\usepackage{booktabs}      
\usepackage{array}         
\usepackage{multirow}      
\usepackage{xcolor}        
\usepackage{colortbl}      
\usepackage{pdflscape}     
\usepackage{adjustbox}     

\usepackage{caption}
\usepackage{algorithm}
\usepackage{algpseudocode}
\usepackage{amsmath}
\usepackage{amssymb}

\theoremstyle{remark}

\theoremstyle{thmstyleone}%
\theoremstyle{thmstyletwo}%

\theoremstyle{thmstylethree}%

\begin{document}
\title[Article Title]{Synapse-Inspired Energy Networks: A Neuromorphic Approach to Microgrid Protection without Communication Links}



\author*[1]{\fnm{Saurabh} \sur{Prabhakar}}\email{eez228471@iitd.ac.in}


\author[1]{\fnm{Bijaya Ketan} \sur{Panigrahi}}\email{bkpanigrahi@ee.iitd.ac.in}

\author[2]{\fnm{Frede} \sur{Blaabjerg}}\email{fbl@energy.aau.dk}

\author[2]{\fnm{Subham} \sur{Sahoo}}\email{sssa@energy.aau.dk}

\affil*[1]{\orgdiv{Department of Electrical Engineering}, \orgname{IIT Delhi},  \city{Hauz Khas}, \postcode{110016}, \state{Delhi}, \country{India}}

\affil[2]{\orgdiv{Department of Energy}, \orgname{Aalborg University}, \city{9220 Aalborg East}, \country{Denmark}}


\abstract{Traditional protection systems for microgrids, which rely on high fault currents and continuous communication, struggle to keep up with the changing dynamics and cybersecurity concerns of decentralized networks. In this study, we introduce a novel biologically inspired protection system based on neuromorphic principles, where each distributed energy resource (DER) functions as a simple neuron. These "neurons" process local changes in voltage, current signals, and converting them into spike patterns that represent the severity of disturbances. Just as neurons communicate via synapses in biological systems, we exploit transmission cables to coordinate between DERs, enabling them to share information and respond to faults collectively. Fault detection and circuit breaker activation are driven by a First-To-Spike (FTTS) mechanism, similar to the concept of traveling wave protection, but without needing GPS synchronization or communication links. A key innovation is the ability to use the timing of spikes to locally determine the nature of a fault, offering an intelligent, adaptive response to disturbances. Performance shows tripping latency of 10--58~ms, surpassing conventional relays and even traveling-wave methods (60~ms), while maintaining detection accuracy above 98\% and spatial selectivity over 97\%, enabling real-time, communication-free, scalable protection for plug-and-play microgrids.}




\maketitle

\section{Introduction}\label{sec1}

As modern power systems evolve towards decentralization, microgrids have become essential for integrating renewable energy and enhancing local resilience \cite{10505017,10526379}. These microgrids are typically powered by inverter-based distributed energy resources (DERs), such as solar PV and energy storage, that offer flexible control but depart fundamentally from traditional protection assumptions \cite{9528967,10399893,altaf2022microgrid}. Unlike synchronous machines, inverters limit fault current based on internal constraints, making it difficult to distinguish faults from routine disturbances \cite{9347814}. Moreover, frequent changes in topology due to DER switching, re-islanding, or varying loads undermine the effectiveness of time-grading and impedance-based protection strategies. Efforts to adapt classical schemes such as directional, distance, and differential relays have introduced enhancements like voltage supervision \cite{10443544}, sequence-droop compensation \cite{10684365}, and adaptive thresholds. However, these technologies still depend heavily on high fault currents or synchronized phasor data, which may be unreliable in inverter-dominated environments \cite{9611029,alasali2024recent}.

To overcome these challenges, recent architectures have embraced communication-assisted strategies. Centralized protection schemes using artificial neural networks~\cite{10387437}, phasor measurement units (PMU)-enabled distributed consensus~\cite{10457952}, and hybrid logic-based protection~\cite{9882102} improve response speed and fault localization. However, these solutions introduce new vulnerabilities, including latency, data loss, jitter, and cyber threats, especially under adversarial conditions.
As a partial remedy, co-transfer communication methods like Power Talk and Talkative Power embed control signals into voltage waveforms or converter dynamics~\cite{he2020nature}. However, they face practical issues including limited bandwidth, sensitivity to EMI, and poor transformer compatibility---constraining their scalability. Semantic communication frameworks that focus on event-driven signaling and predictive timing~\cite{10485192} offer additional resilience but remain complex to implement.

At the same time, travelling wave (TW) protection has gained attention as a communication-free method that detects faults within a fraction of a cycle by sensing high-frequency wavefronts generated during a fault event. TW schemes, which use GPS-synchronized sensors, work well in large or hybrid power systems~\cite{entsoe2021tw}. However, their effectiveness decreases in low-voltage or meshed microgrids, where signal reflections, noise, and weak waveforms from nearby faults can interfere with detection. These challenges highlight the need for new protection strategies that are ultra-fast, locally operated, and do not rely on communication—particularly for low-inertia and reconfigurable microgrids.

Neuromorphic protection is an emerging approach that applies principles from biological neural systems to power networks, enabling local, event-driven intelligence by leveraging the physical wiring between sources in a microgrid~\cite{10839209,song2025spike}. The \textit{Spike Talk} architecture was one of the first to introduce this concept to power systems, modeling each distributed energy resource (DER) as a leaky integrate-and-fire (LIF) neuron. These neurons autonomously monitor local voltage and current signals and generate spike trains to encode operational disturbances. Unlike conventional systems that rely on continuous communication, these spikes are processed asynchronously and locally, with responses triggered only when signal thresholds are exceeded--allowing for decentralized and real-time fault detection~\cite{10909727}.

Building on this foundation, Diao et al.~\cite{diao2024inferring, 10445758} proposed event-driven coordination frameworks using spiking neural networks (SNNs), in which decisions are initiated by meaningful events rather than continuous data exchange. This represents a conceptual shift—from systems that communicate about power to systems that \textit{reason} with power—where protection decisions are guided by spike timing and the internal dynamics of the network itself.

Notably, the spike-based mechanism in neuromorphic protection parallels the logic of traveling wave (TW) protection schemes. TW methods detect and localize faults by identifying the earliest arrival of high-frequency wavefronts at spatially distributed sensors. Similarly, in the neuromorphic approach, the first DER to spike occurrence indicates the fault’s proximity. However, unlike TW systems, which require GPS synchronization, high-speed sensors, and dedicated communication infrastructure, the neuromorphic method operates without any of these dependencies. Instead, it provides a scalable, and inherently communication-free solution for fast and resilient fault detection.

To address the limitations of conventional protection systems, we introduce a neuromorphic framework in this paper inspired by the dynamics of biological neurons. In this approach, each distributed energy resource (DER) is modeled as a leaky integrate-and-fire (LIF) neuron that continuously monitors local voltage and current. These neurons emit spikes in response to disturbances, with larger or closer faults causing earlier spikes. The DERs that spikes first extrapolates to be nearest to the fault and autonomously initiates selective circuit breaker tripping, thereby enabling ultra-fast, fully decentralized protection without the need for communication or synchronization.

This First-to-Spike (FTTS) logic mirrors competitive neural behavior in biological systems, where the earliest spike encodes the most critical stimulus. Notably, the relationship between spike timing and disturbance magnitude follows an inverse-time characteristic, conceptually similar to the inverse definite minimum time (IDMT) curves used in traditional protection relays. This parallel establishes a mathematical and functional duality between neuromorphic encoding and classical relay logic. Figure 1 summarizes the overall architecture, illustrating how LIF neuron models, inverse-time spike responses, and local breaker actuation are combined to realize a biologically-inspired protection system for microgrids.
\begin{figure}
    \centering
    \includegraphics[scale=0.95]{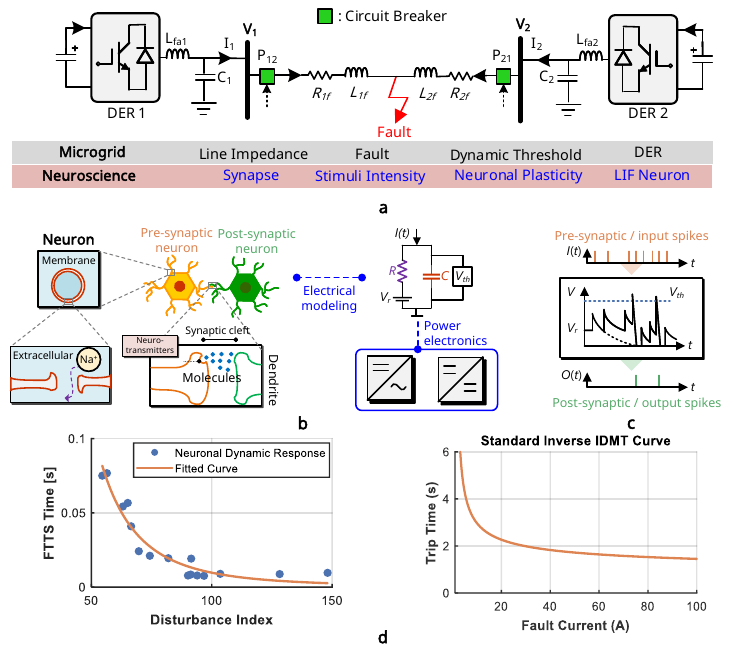}
    \caption{\textbf{Neuromorphic foundation of spike-based protection for AC microgrids}:
(a) A two-bus AC microgrid setup with inverter-based DERs connected via a tie-line, protected by circuit breakers at both ends. Each DER functions as an independent spiking unit, allowing fully decentralized and communication-free protection. (b) Each DER is modeled as a Leaky Integrate-and-Fire (LIF) neuron, where inverter-side capacitance and line resistance together replicate the behavior of a biological neuron's membrane using an equivalent RC circuit. (c) The neuron accumulates input spikes over time, and once the membrane potential crosses a voltage threshold, it emits an output spike—mimicking how biological neurons fire. (d) Increasing disturbance severity shortens the spike interval, resulting in faster spike emission. This inverse-latency characteristic is mathematically equivalent to the IEC 60255 Standard Inverse time curve, establishing a direct analytical bridge between classical IDMT protection and the proposed neuromorphic encoding.}

    \label{fig:control_schematic}
    \vspace{-2mm}
\end{figure}

\begin{figure}
    \centering
    \includegraphics[scale=0.90]{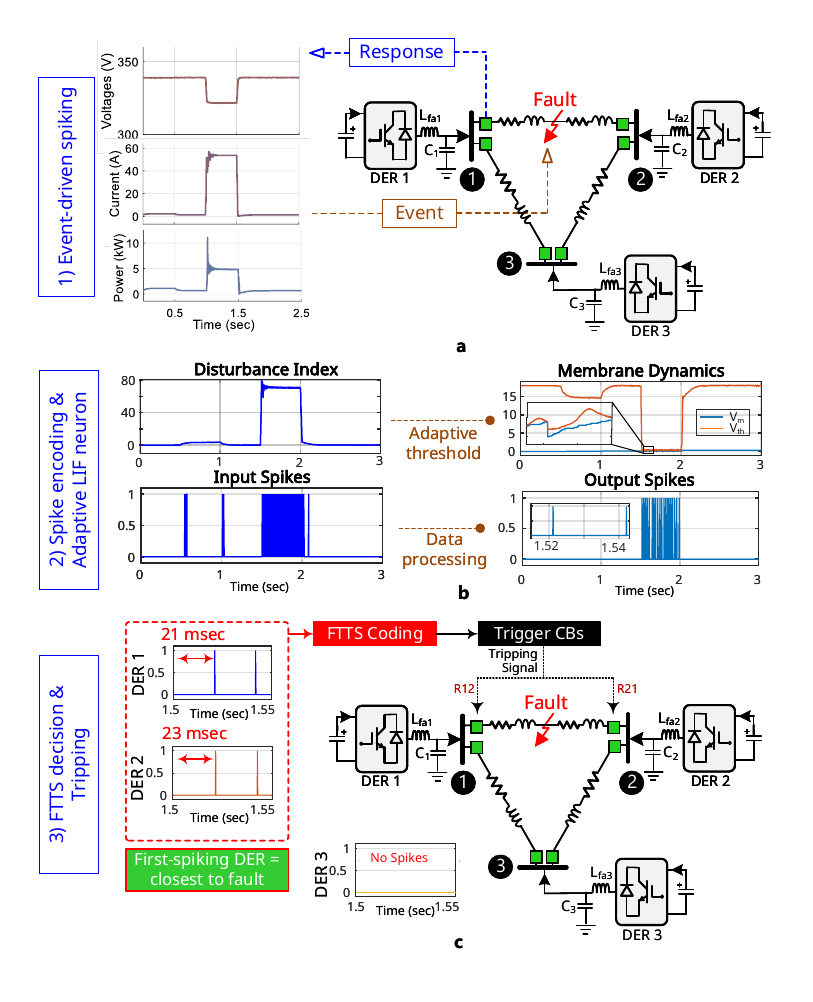}
    \caption{\textbf{Schema of the proposed spike-based neuromorphic protection framework}, which begins with (a) event-driven sensing, where a load change around 0.5\,s and a fault near 1.5\,s create observable changes in voltage, current, and power—resulting in a rise in the disturbance index $D(t)$. These signals are then passed through (b) spike encoding, where local deviations are converted into input spike trains based on their severity. In membrane integration and adaptive thresholding, the neuron integrates the input spikes, and the membrane potential $V_{\mathrm{m}}(t)$ rises until it crosses the adaptive threshold $V_{\mathrm{th}}(t)$, triggering an output spike. Finally, (c) the first-spiking neuron sends a trip command to its local circuit breaker, isolating the faulted line. The entire response is fast, local, and communication-free decision-making through asynchronous spiking dynamics.}

    \label{fig:control_schematic}
    \vspace{-2mm}
\end{figure}

To implement this architecture in practice, each DER locally computes a scalar disturbance index defined as: 
\begin{equation}
    D(t) = \alpha |\Delta V(t)| + \beta |\Delta I(t)| + \gamma |\Delta P(t)|
\end{equation} 
where, the coefficients $\alpha$, $\beta$, and $\gamma$ weigh voltage, current, and power deviations respectively. This index modulates the inter-spike interval via an inverse-time relationship: $T_{s}(t) = \frac{1}{1 + k D(t)}$, where $k > 0$ governs the sensitivity to disturbances. As fault severity increases, the disturbance index $D(t)$ rises, resulting in shorter inter-spike intervals and earlier spike generation. This mechanism effectively emulates the inverse definite minimum time (IDMT) characteristics of conventional protection relays, but through a biologically inspired, event-driven process.

Conceptually, this timing-based response mirrors the principles of traveling wave (TW) protection, where the earliest arrival of a wavefront indicates the fault’s location. In the proposed framework, shorter inter-spike intervals similarly correspond to higher local disturbances, enabling spatially selective fault localization based on spike latency—without requiring GPS synchronization or high-frequency sensing.
To enhance robustness during normal operation, an adaptive thresholding mechanism is incorporated. This threshold evolves dynamically, informed by both the real-time disturbance index and the membrane potential of the LIF neuron. Consequently, spikes are generated only in response to statistically significant anomalies, effectively suppressing false positives arising from routine load variations or minor transients.

The ensemble of spike trains generated across DERs forms an asynchronous decision layer, in which the first neuron to spike autonomously activates its associated circuit breaker (CB). This \textit{First-to-Spike} (FTTS) mechanism\footnote{It should be noted that First-to-Spike coding mechanism can normally be abbreviated to FTS and FTTS. We use FTTS consistently throughout this work.} \cite{proc} enables sub-cycle fault protection based entirely on local sensing and spike timing, without the need for communication or synchronization. By eliminating centralized coordination, the system enhances resilience to cyber-physical threats, communication delays, and network failures.This marks a shift from conventional magnitude- or phasor-based logic to a fundamentally new protection paradigm where decisions arise from spike-timing computation.

Figure 2 illustrates the complete neuromorphic protection workflow—from local disturbance sensing to spike generation and CB actuation—highlighting the architecture’s decentralized, biologically inspired logic. \textit{Supplementary Fig. 1} further contrasts this approach with conventional hierarchical microgrid protection schemes, which typically rely on intelligent electronic devices (IEDs), central controllers, and continuous communication. In contrast, the proposed neuromorphic design achieves fast, robust, and fully localized decision-making.Notably, the membrane dynamics of the LIF neuron naturally produce an inverse-time relationship between disturbance severity and spike latency, establishing an emergent analogue to classical IDMT behavior without requiring preset curves.

To evaluate the proposed framework, we performed extensive simulations across diverse fault and impedance conditions. The neuromorphic system achieved sub-cycle detection and isolation within 10--58 ms: ABCG faults were cleared in 10 ms, ABG in 28 ms, and AG in 58 ms, compared with the 200--450 ms response times of conventional microgrid protection. Spike latencies showed a strong correlation with fault severity, reflecting inverse-time characteristics but realized through a local, event-driven architecture. Unlike traveling-wave schemes that require GPS synchronization and high-speed sensors for 60 ms operation, the neuromorphic approach achieves faster and fully selective protection using only local spike timing, making it ideally suited for scalable, plug-and-play microgrids \cite{sel_traveling_wave}.

In addition to speed and accuracy, the framework offers several operational advantages. The adaptive thresholding mechanism reliably suppresses false positives during normal load variations, while the First-to-Spike logic ensures that faults are localized by the DER nearest to the disturbance. This enables effective protection even in meshed networks with redundant paths. Overall, the proposed approach bridges classical protection theory with neuromorphic computing. By combining spike-timing computation, emergent inverse-time behavior, and real-time validation on the OP4500 hardware-in-the-loop platform, it introduces a biologically inspired paradigm for decentralized power system protection—one that senses locally, reacts instantly, and adapts naturally to the evolving dynamics of distributed energy networks.

\section{Results}\label{sec2}

We assess the proposed spike-based neuromorphic protection framework in terms of speed, accuracy, and scalability across microgrids of varying sizes and disturbance conditions. In this approach, each distributed energy resource (DER) is modeled as a leaky integrate-and-fire (LIF) neuron that continuously monitors local voltage, current, and power. Disturbances are encoded into spike timings, with more severe events triggering faster spiking activity. Protection decisions follow a \textit{First-to-Spike} (FTTS) rule: the DER that spikes first initiates local circuit breaker actuation. We evaluate the system under a range of conditions, including single and multiple fault events, as well as benign load variations, to demonstrate its robustness and adaptability in realistic, dynamic microgrid environments.

\subsection{Spike-Based Fault Response Across Different Network Conditions}

\subsubsection*{Case 1: {Fault Detection Without Breaker Action}}

To isolate and examine detection behavior, we first study a three-bus ring microgrid (Supplementary Fig. 2a) with all circuit breakers disabled. As shown in Supplementary Fig. 3, a normal load increase occurring at 0.5 s results in a moderate rise in the disturbance index. The adaptive threshold responds gradually to this change, adjusting in line with the disturbance and effectively preventing the generation of unnecessary spikes. This behavior demonstrates the system’s ability to maintain selectivity and avoid false positives during routine operational variations. Later in the simulation, when a single-line-to-ground (SLG) fault occurs, the disturbance index $D(t)$ increases sharply. The spike interval shortens, and the membrane potential $V_m (t)$ rapidly approaches the adaptive threshold, causing affected DERs to fire with sub-60 ms latency of fault onset. This confirms that the framework detects faults quickly and with precise timing.

Since the breakers are disabled, the fault is not isolated and propagates through the network, leading to distorted line currents and frequency deviations. This experiment highlights the neuromorphic system’s ability to provide fast local fault detection through biologically inspired spike-based encoding without GPS synchronization or high-speed sensing.

\subsubsection*{Case 2: Fast Fault Isolation with FTTS-Enabled Breaker Operation in a Ring Microgrid}

We then evaluate the three-bus ring microgrid with the \textit{First-To-Spike} (FTTS) protection logic enabled. Supplementary Fig. 4 demonstrates that when a fault occurs, the distributed energy resource (DER) closest to the fault experiences the highest disturbance index and is the first to generate spike. This early spike autonomously triggers the circuit breakers at both ends of the faulted line. The protection response is completed within approximately two cycles—well before conventional current transients reach their peak—indicating that the decision is governed by spike timing rather than current magnitude. Once the faulty segment is isolated, the rest of the microgrid stabilizes rapidly. These results confirm that fast, directional, and localized protection can be achieved entirely through decentralized, spike-based logic via the wired transmission lines.

\subsubsection*{Case 3: Multi-Fault Adaptation and Scalability in a Meshed Microgrid}

To evaluate adaptability in more complex topologies, we simulate a four-bus meshed microgrid (Supplementary Fig. 2b) subjected to two consecutive disturbances: first, a line-to-line (LL) fault, followed by a more severe three-phase-to-ground (LLLG) fault. As shown in Supplementary Fig. 5, each DER responds independently based on local measurements. The initial LL fault generates moderate spike activity, while the subsequent LLLG fault induces faster and denser spiking due to its higher severity. The adaptive threshold adjusts dynamically, lowering in response to the stronger disturbance and enabling even earlier firing.

In both cases, the corresponding circuit breakers trip within a fraction of a cycle, isolating only the affected lines. The rest of the microgrid continues to operate without disruption. This experiment confirms that the FTTS-based protection framework scales effectively to meshed networks, adapts to varying fault types and magnitudes, and delivers precise, selective fault isolation—without requiring global coordination or communication.

\subsection{Stronger Faults Always Cause Faster Tripping in Spike-Based Protection}

DERs on low-impedance lines detect and trip faster than those on higher-impedance paths. For instance, as shown in Fig.~3a, a DER connected through $z = 2\,\Omega$ trips much earlier than one through $z = 6\,\Omega$, confirming that detection speed is governed by electrical distance, with nearby faults isolated sooner than distant ones.The membrane time constant also plays a role: a smaller constant allows the membrane potential to rise faster, leading to earlier spiking. In Fig.~3b, a neuron with $\tau = 86.4~\mathrm{\mu s}$ trips noticeably faster than one with $\tau = 161~\mathrm{\mu s}$ for the same fault, showing that relay-like time grading can be tuned directly through neuron dynamics.

Fault severity further enhances the selectivity of the protection response. The gain parameter 
$k$ amplifies the impact of disturbances on spike generation, effectively modulating the system’s sensitivity. As shown in Fig. 3c, three-phase faults induce the fastest spiking responses, followed by line-to-line and single-line-to-ground faults. This behavior demonstrates that the spike-based framework naturally prioritizes more severe disturbances, closely mirroring the logic of inverse-time relays.

The combined influence of line impedance, membrane time constant, and fault severity is summarized in Fig. 3d. A three-dimensional surface maps tripping time as a function of disturbance magnitude and fault resistance. The consistently downward slope confirms that more severe and lower-resistance faults lead to faster isolation. These results reinforce that the neuromorphic protection framework replicates the key principles of classical inverse-time protection—achieving fast, graded, and fault-selective responses.

\begin{figure*}
    \centering
    \includegraphics[scale=0.85]{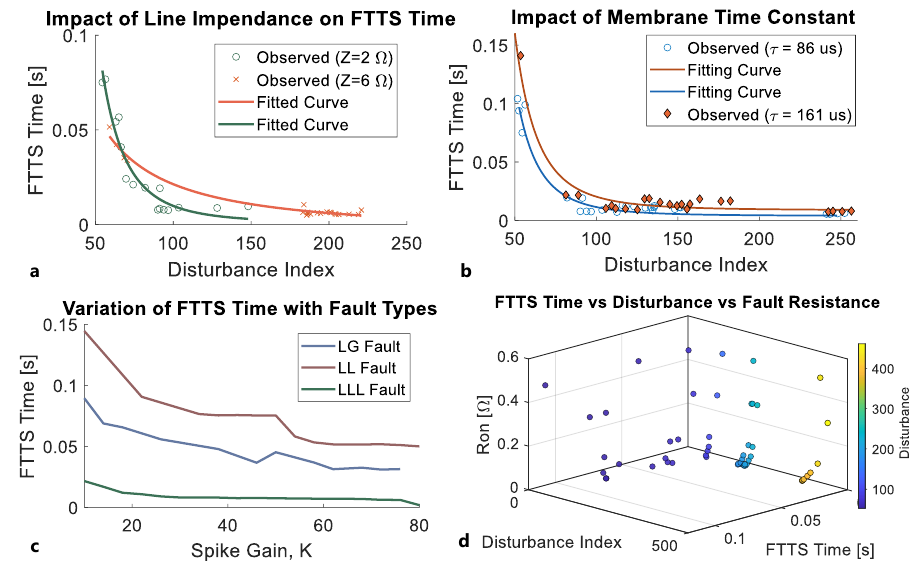}
    
    \caption{\textbf{Influence of electrical and neuromorphic parameters on spike-driven fault response.} (a) DERs connected through lower impedance lines (Z=$2\,\Omega$) detect faults faster than those on higher impedance lines (Z=$6\,\Omega$), as stronger disturbances lead to earlier spikes. The observed inverse-time behavior matches well with classical relay logic. (b) A neuron with a smaller membrane time constant ($\tau = 86\,\mu s$) spikes earlier than one with larger $\tau = 161\,\mu s$, highlighting how spike timing can be tuned biologically to reflect fault location. (c) Tripping time varies with fault severity---quickest for three-phase faults (LLL), slower for line-to-line (LL), and slowest for single-line-to-ground (LG)---mimicking classical selectivity. (d) The 3D surface shows how tripping time decreases with increased disturbance and lower fault resistance, replicating the behavior of inverse-definite minimum time (IDMT) curves using a fully local, neuromorphic logic.}
    \label{fig:control_schematic}
    \vspace{-2mm}
\end{figure*}

\subsection{Accuracy and Selectivity of Spike-Based Protection}

We tested the spike-based protection framework under a wide range of scenarios, including variations in fault resistance and sudden load changes. As shown in Fig.~4a, the framework provides fast and reliable fault detection while avoiding false trips during normal fluctuations. The membrane time constant $\tau$ strongly influences performance: when $\tau$ is small ($\tau = 86~\mu\mathrm{s}$), the neuron integrates faster, fires earlier, and achieves over 99\% detection accuracy. In contrast, a larger value of $\tau$ (e.g., $\tau = 161~\mu\mathrm{s}$) slows the membrane response, lowering accuracy to approximately 81\%. Even under load variations, the adaptive threshold mechanism prevents spurious spiking, resulting in 100\% correctness. Across all tested cases, the system maintains an average fault detection accuracy above 98\%.

Fault resistance also plays a critical role. As resistance increases, the magnitude of the disturbance signal decreases, making some fault events more difficult to detect. As shown in Fig.~4b, the framework continues to identify most faults correctly. Some weak single-line-to-ground faults at higher resistance values may escape detection; however, the system still maintains selectivity by tripping only the affected node, thereby avoiding unnecessary isolation of healthy parts of the network.

Together, these results demonstrate that spike timing naturally adapts to fault strength: stronger disturbances induce earlier spikes, while normal load changes are effectively ignored. This dynamic balance allows the framework to deliver fast, selective, and communication-free protection suitable for microgrid ecosystem.

\begin{figure*}
    \centering
    \includegraphics[scale=0.85]{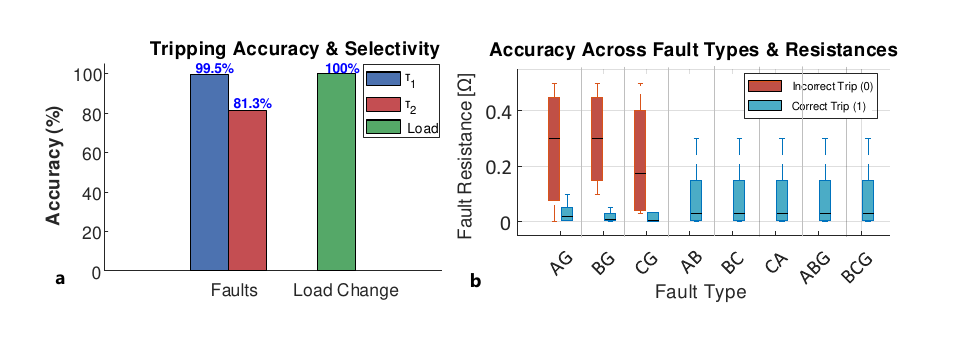}
    
    \caption{\textbf{Accuracy and selectivity trends in spike-based neuromorphic protection:} 
(a) Grouped bar plot illustrating the tripping accuracy of the spike-based neuromorphic protection scheme across different operating scenarios, including low and high membrane time constants and load disturbance cases. The framework maintains high accuracy ($>99\%$) under low $\tau$, around $81.3\%$ under high $\tau$, and $100\%$ under load disturbances, highlighting robustness and selectivity.(b) Box plot showing tripping accuracy across varying fault resistances and fault types. While the system performs reliably up to $0.5~\Omega$, occasional misclassifications occur in LG faults due to weaker disturbance signatures. The results affirm the role of spike timing in capturing fault severity and supporting accurate, selective protection decisions.
}
    \label{fig:control_schematic}
    \vspace{-2mm}
\end{figure*}

\subsection{Energy Efficiency, Selectivity, and Scalability in Spike-Based Protection}

The proposed neuromorphic framework shows two essential properties: it remains almost silent during normal operating conditions and becomes highly selective during faults, enabling fast and localized isolation. During load changes in microgrid, the system conserves energy by suppressing unnecessary spikes. As seen in Fig.~5a--b, the disturbance index and spike activity rise only slightly, leading to just 5 spikes for a 20\% increase in load and 11 spikes for a 40\% increase. This quiet behavior demonstrates minimal energy use, as the adaptive threshold adjusts smoothly to block false alarms.

When faults occur, the response becomes both urgent and proportional to severity. In Fig.~5b, the membrane potential $V_m$ exceeds the adaptive threshold $V_{th}$ only during critical disturbances, producing fast and dense spike trains. For AG, ABG, and ABCG faults, the number of spikes rises sharply to 477, 688, and 831 in Fig. 5c, with the first spikes (corresponding to the fault) appearing at 58~ms, 28~ms, and 10~ms, respectively. This pattern mirrors the inverse-time principle of conventional relays, where stronger faults are always cleared faster.

Under the \textit{First-To-Spike} (FTTS) rule, only the distributed energy resource (DER) closest to the fault initiates spiking, while all others remain inactive. As shown in Fig.~5d, this ensures spatial selectivity: only the affected section is isolated, while the rest of the microgrid continues to operate undisturbed. To further aid intuition, Supplementary Fig.~11 provides a clear system-level view, showing how spike-based protection adapts seamlessly to both fault severity and fault location across all DERs. 

Our framework is designed to be scalable, as demonstrated by the test system results across four different configurations (Supplementary Figs. 2, 8). The effectiveness of this approach in varying system sizes and fault scenarios is further validated by the results in Supplementary Figs. 4, 5, 9, and 10, where it is shown that the spike-based protection mechanism adapts seamlessly to both fault severity and fault location. This scalability ensures that our solution remains robust and efficient regardless of system complexity or fault conditions.

This behavior arises naturally from the underlying LIF neuron dynamics. Small, gradual disturbances during normal operation cause slow membrane potential changes that rarely exceed the adaptive threshold, producing minimal spikes. In contrast, sudden, severe faults rapidly increase the disturbance index, causing the membrane potential to cross the dynamically lowered threshold and trigger immediate spikes. The spike interval compresses as the disturbance magnitude grows, offering a direct, real-time indication of fault severity. Importantly, this mechanism remains effective under current-limited operation of grid-forming (droop-controlled) DERs, where output currents saturate rapidly during faults, because the disturbance index is jointly driven by voltage, current, and active-power deviations rather than current alone. The influence of DER-side overcurrent limiting on spike generation, membrane dynamics, and FTTS decision-making is explicitly analyzed and validated in Supplementary Note 13. This mechanism allows the system to achieve both energy-efficient monitoring and fault-proportional responses without explicit communication or centralized coordination.Together, these properties highlight how energy efficiency, fast response, and spatial selectivity emerge intrinsically from the spike-based, neuromorphic protection paradigm, providing a foundation for robust, autonomous microgrid protection.

\begin{figure*}
    \centering
    \includegraphics[scale=1]{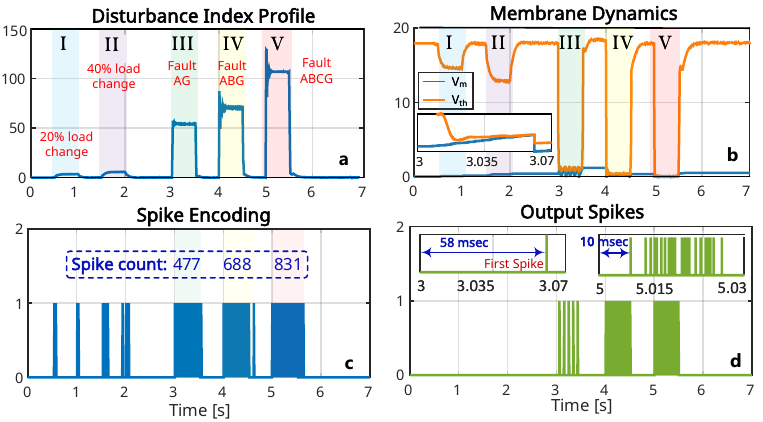}
    
  \caption{Energy-efficient and fault-sensitive spike response. 
The neuromorphic protection system shows selective activity based on disturbance severity. 
In (a), the disturbance index $D_i(t)$ rises gradually during load changes but shows a sharp increase during fault events. 
This is mirrored in (c), where the input spike activity remains sparse under load shifts but becomes dense and urgent under fault conditions. 
In (b), the membrane potential $V_\text{m}(t)$ increases over time and surpasses the adaptive threshold $V_\text{th}(t)$ only during critical disturbances, triggering output spikes. 
Subplot (d) highlights the timing of these output spikes, showing that more severe faults trigger faster responses—58~ms for an AG fault and just 10~ms for an ABCG fault. 
Finally, (b) shows that the total number of spikes scales with fault severity: 477 spikes for AG, 688 for ABG, and 831 for ABCG, demonstrating how the system maintains energy-efficient behavior during normal events while intensifying its response during critical faults.}

    \label{fig:control_schematic}
    \vspace{-2mm}
\end{figure*}

\section{Discussion}

As microgrids evolve into decentralized systems dominated by inverter-based sources, conventional protection strategies---rooted in centralized coordination and high fault currents---face increasing limitations. This study introduces a biologically inspired neuromorphic protection framework, where each Distributed Energy Resource (DER) functions as a Leaky Integrate-and-Fire (LIF) neuron, locally encoding disturbances in voltage, current, and power into asynchronous spike trains. Protection decisions follow a First-To-Spike (FTTS) logic, analogous to traveling-wave methods that rely on the earliest signal arrivals for fault localization, but achieved here without requiring communication, GPS synchronization, or centralized coordination.

Simulation results show that spike timing dynamically mirrors the inverse-definite minimum time (IDMT) principle: stronger faults induce faster spikes and earlier trips, achieved through membrane integration rather than preset relay curves. An adaptive thresholding mechanism further ensures reliability, suppressing benign events while permitting rapid action during genuine faults. Critically, only the DER electrically closest to the fault spikes first, enabling local isolation without disrupting the wider network.

Beyond accuracy and speed, the framework is also energy- and computation-efficient. As shown in Section~2.4, the system remains dormant during steady-state conditions and activates only in response to significant disturbances. Spike count and computational effort scale with fault severity—reflecting the brain’s efficiency in responding only to meaningful stimuli. More than just emulating conventional protection, this architecture extends and redefines it. Its spike-timing behavior is mathematically equivalent to IDMT relays, yet the decentralized, event-driven nature makes it scalable and plug-and-play. New DERs can be added seamlessly, without system reconfiguration, making it ideal for dynamic, peer-to-peer microgrids.

Looking ahead, this framework lays the foundation for adaptive and learning-based protection. Integrating mechanisms like \emph{Spike-Timing Dependent Plasticity} (STDP) \cite{proc} could enable real-time threshold adaptation and context-aware responses. Beyond fault detection, the proposed biologically inspired protection framework opens a broad landscape for future research. Because disturbances are encoded as sparse, event-driven spike trains rather than dense, continuous waveforms, the method inherently filters out quiescent periods and concentrates computation only around informative transients, which reduces data volume and processing burden. This same spike-based representation can be reused for complementary protection functions such as islanding detection, event classification, and fault-type localization by exploiting the spatio-temporal firing patterns that emerge across multiple neuron-like DER agents.The formulation also naturally supports adaptive learning mechanisms—such as spike-timing-dependent plasticity or local rule-based threshold adaptation—which could enable automatic tuning under ageing, topology reconfiguration, or evolving operating conditions. Furthermore, the sparse and event-driven nature of spike processing makes the approach well suited for implementation on emerging neuromorphic hardware platforms, offering the potential for sub-millisecond protection decisions at extremely low computational cost. Together, these directions strengthen the broader vision of neuromorphic protection as a foundation for intelligent, autonomous, and resilient next-generation microgrids.

Although the proposed neuromorphic protection framework is fully decentralized and lightweight, several practical considerations must be acknowledged for real-world deployment. First, scalability across larger microgrids is inherently supported because each DER neuron operates locally without communication; however, extremely large networks may require additional filtering to avoid spike collisions when many nodes experience simultaneous but mild disturbances. Moreover, while the core studies employ identical DER control structures for clarity, the framework is not restricted to homogeneous units; its applicability to microgrids with heterogeneous DER ratings and control dynamics has been explicitly examined, with detailed modeling, parameter sets, performance results, and discussion provided in Supplementary Tables 5–6, Supplementary Figs. 16–17, and Supplementary Note 12. Second, the method remains robust under weak-waveform or low-voltage faults—conditions where TW relays typically struggle—because spike generation depends on aggregated voltage, current, and power deviations rather than high-frequency components. Even when the waveform is distorted or the voltage sag is shallow, the disturbance index still grows sufficiently to compress spike intervals and trigger a reliable FTTS response. Third, the framework assumes availability of accurate local sensing at each DER and a digital controller capable of executing simple LIF updates at microsecond resolution, both of which are standard in present-day inverter hardware.Nonetheless, the method has limitations. Extremely high-resistance faults may produce disturbances just above the adaptive threshold, leading to slower spikes compared to low-impedance faults. Similarly, if extensive DER clustering results in nearly identical disturbance levels across nodes, additional tie-line impedance information may be required to ensure unique FTTS resolution. These considerations do not undermine the method’s applicability but highlight realistic boundaries and avenues for refinement. Overall, while fully compatible with a wide range of topologies and operating conditions, the neuromorphic approach—like all protection schemes—benefits from careful parameter tuning to ensure robust operation under extreme or highly ambiguous scenarios.


\section{Methods}

\subsection{System Setup and Test Scenarios}

We developed a low-voltage islanded AC microgrid model in MATLAB/Simulink (R2023a) to evaluate the proposed neuromorphic protection strategy. The grid operated at {415} {V}, {50} {Hz}, and included inverter-based DERs, each rated at {10} {kW}. Two representative configurations were tested: a radial network, where power flows along a single path, and a meshed network, characterized by multiple parallel paths that increase complexity in fault detection and isolation.

Distribution lines were parameterized with resistance and reactance values, capturing characteristics of both underground and overhead cables. Detailed line parameters are provided in \textit{Supplementary Tables~1 and 2}. Each DER was equipped with local current and voltage sensors, enabling protection decisions to be made solely from local measurements without communication or centralized coordination. Two disturbance types were applied. First, load variations of 20\% and 40\% simulated normal demand fluctuations. Second, faults were injected at various locations, including single-line-to-ground (AG), line-to-line (AB), and three-phase (ABC) faults. Fault resistance ranged from {0.001} $\Omega$ to {3} $\Omega$, encompassing solid short circuits and high-impedance faults such as insulation breakdown or arcing.

Over 300 simulation runs were conducted across both network topologies, encompassing symmetrical and asymmetrical faults, a broad impedance spectrum, and multiple DER placements. These scenarios enabled evaluation of four critical metrics: detection speed, selectivity, resilience to false trips, and performance under different network configurations. Crucially, the neuromorphic protection relied exclusively on local sensing and spike-based neuron dynamics.

\subsection{Neuromorphic Modeling of DERs as Leaky Integrate-and-Fire (LIF) Neurons}

Each inverter-based DER was modeled as a leaky integrate-and-fire (LIF) neuron, converting local disturbances into spike signals. This biologically inspired design empowers each DER to independently make protection decisions without external coordination. To detect disturbances in the microgrid, each DER continuously compares its measured electrical quantities with their steady-state (pre-fault) values, denoted as \( V_0 \), \( I_0 \), and \( P_0 \). The deviations are calculated as:
\begin{equation}
\Delta V(t) = \left| V(t) - V_0 \right|,\quad 
\Delta I(t) = \left| I(t) - I_0 \right|,\quad 
\Delta P(t) = \left| P(t) - P_0 \right|
\end{equation}
The disturbance index at any DER is defined as (1):

\[
D(t) = \alpha \left| \Delta V(t) \right| + \beta \left| \Delta I(t) \right| + \gamma \left| \Delta P(t) \right|
\]

where \( \Delta V(t) \), \( \Delta I(t) \), and \( \Delta P(t) \) represent the instantaneous deviations of voltage, current, and active power from their steady-state values. Here, the voltage and current quantities correspond to aggregated local three-phase RMS measures, while the active-power term represents the instantaneous three-phase total active power computed from the sum of per-phase contributions. This unified representation is adopted for both symmetrical and asymmetrical fault conditions. Under unbalanced faults, phase-level asymmetries, negative-sequence components, and second-order power oscillations naturally manifest as increases in these aggregated quantities, thereby producing nonzero deviations in all three channels. As a result, the disturbance index inherently captures the presence and severity of asymmetrical faults without requiring explicit per-phase processing or sequence decomposition, preserving the simplicity and locality of the proposed formulation.
The weighting coefficients are \( \alpha = 1.0 \), \( \beta = 0.5 \), and \( \gamma = 0.005 \), balancing contributions from all three signals.The chosen coefficients ($\alpha = 1.0$, $\beta = 0.5$, $\gamma = 0.005$) follow a scaling rationale based on the relative magnitudes of the underlying signals. In inverter-dominated microgrids, $\Delta V$ and $\Delta I$ typically remain within $0$--$0.3\,\text{pu}$, whereas the active-power deviation $\Delta P$ can be numerically one to two orders larger. To prevent $\Delta P$ from dominating the disturbance index, $\gamma$ is selected several orders smaller, while $\alpha$ and $\beta$ are chosen to ensure a balanced contribution of the voltage and current components.

In Supplementary Fig.~7, a detailed sensitivity analysis was conducted to evaluate robustness with respect to the disturbance-index weights. A parameter sweep of $\alpha \in [0.5,\,2.0]$, $\beta \in [0.25,\,1.5]$, and $\gamma \in [0.001,\,0.02]$ was performed over 400 fault scenarios. Across these ranges, the detection accuracy remained $\ge 93.5\%$, and the median tripping latency varied by less than $8\,\text{ms}$, indicating a broad plateau of stable performance around the nominal set $(\alpha,\beta,\gamma)$.

The disturbance index drives the neuron’s membrane potential \( V_m(t) \) according to the differential equation:

\begin{equation}
C_m \frac{dV_m(t)}{dt} = -\frac{V_m(t)}{R_m} + D_i(t)
\end{equation}

where \( C_m = 250\,\mu\text{F} \) and \( R_m = 0.644\,\Omega \), yielding a response time constant \( \tau = C_m R_m = 162.5\,\mu\text{s} \). Small disturbances decay naturally, while significant ones cause rapid potential rises. A spike is generated when \( V_m(t) \) exceeds an adaptive threshold:

\begin{equation}
V_{\text{th}}(t) = V_0 \exp\left[ -\eta\left( |\Delta V| + |\Delta I| + |\Delta P| \right) \right] + \lambda V_m(t)
\end{equation}

with \( V_0 = 17.9\,\text{V} \), \( \eta = 0.008 \), and \( \lambda = 0.6 \). This adaptive threshold ensures sensitivity to genuine faults while filtering noise. Complete LIF neuron parameters, including variations to test robustness, are listed in \textit{Supplementary Table 3}. After spiking, the inter-spike interval \(T_{s}(t)\) depends on disturbance severity:

\begin{equation}
    T_{s}(t) = \frac{1}{1 + k D(t)}
\end{equation}

where, the gain \(k\) modulates spiking speed: higher \(k\) leads to faster spikes and quicker breaker response, whereas lower \(k\) results in more conservative operation. Thus, each DER functions as an event-driven, self-contained neuron, linking local disturbances directly to breaker trips.

\subsection{First-to-Spike (FTTS) Tripping Mechanism}

Protection decisions utilize a First-To-Spike (FTTS) rule among all DER neurons: the one generating the earliest spike is assumed closest to the fault and immediately trips its breaker. This guarantees rapid, decentralized, and selective fault isolation. Unlike phasor-based methods relying on communication and global thresholds, FTTS is inherently event-driven. Only one breaker operates per fault event, namely the one associated with the first spiking DER. Near-simultaneous spikes (within {3.125} {$\mu s$}) are resolved via ''\textit{a winner-takes-all}'' mechanism that suppresses subsequent responses.

Across more than 300 fault scenarios—including meshed topologies and high-impedance conditions—the FTTS rule achieved overall exceeding 98\% location accuracy and an average detection time below 25 ms, demonstrating robustness against both strong and weak fault currents. Consequently, the neuromorphic strategy achieved fast, selective, and communication-free protection, a critical advantage for inverter-dominated microgrids where traditional schemes often fail.

\subsection{Simulation Setup and Data Logging}

All simulations were performed in \textsc{MATLAB}/Simulink (R2023a) using a fixed-step solver with an integration step of \( 3.125\,\mu\text{s} \), consistent with the LIF neuron dynamics. At each step, key variables were logged, including membrane potential \( V_m(t) \), spike output \( S(t) \), disturbance index \( D(t) \), adaptive threshold \( V_{\text{th}}(t) \), breaker status, and the three-phase voltage, current, and active power at each DER. Fault initiation and clearing times were also recorded to evaluate protection delays and accuracy.

The neuromorphic LIF neuron parameters used to model DER units are summarized in \textit{Supplementary Table~3.} Variations across neurons were introduced to study parameter sensitivity, providing a comprehensive view of the system’s spike-based response under different disturbance conditions. Spike outputs were stored as binary event sequences and post-processed to extract inter-spike intervals, spike counts, and breaker activation times. Visualization plots, such as spike rasters, membrane potential traces, and breaker timing graphs, were generated using \textsc{MATLAB}’s plotting tools. This structured pipeline ensures full reproducibility and systematic benchmarking of the neuromorphic protection framework.

\subsection{Metric Extraction: \textit{Latency}, \textit{Accuracy}, and \textit{Spike Economy}}

To evaluate the neuromorphic protection system, we defined three key performance measures: \textit{tripping latency}, \textit{detection accuracy}, and \textit{spike economy}. All metrics were extracted from \textsc{MATLAB}/Simulink simulations using automated scripts across more than 300 test cases.

\subsubsection{Tripping Latency}

Tripping latency quantifies the system’s speed of response. It is measured as the time difference between the onset of a fault and the first spike that triggers a breaker:
\begin{equation}
\text{Latency} = t_{\text{spike}} - t_{\text{fault onset}}
\end{equation}

The fault onset is marked when current exceeds 5\% of its pre-fault steady-state level. The 5\% value is determined from extensive empirical analysis of fault and load-disturbance scenarios. Across all tested conditions, large load variations (up to 40\%) resulted in current deviations of only 1--3\%, whereas high-resistance faults consistently produced deviations of $\ge 5\%$, with solid faults exceeding 20\%. Therefore, 5\% represents the smallest deviation that reliably separates normal operating variations from genuine faults in our study.This metric allows comparison of reaction speed across different fault types, resistances, and locations.

\subsubsection{Detection Accuracy}

Detection accuracy assesses the correctness of fault identification. A case is considered accurate if the DER closest to the fault spikes within two grid cycles (60~ms at 50~Hz), and its breaker clears the corresponding line. Accuracy is expressed as:
\begin{equation}
\text{Accuracy} = \frac{N_{\text{correct}}}{N_{\text{total}}} \times 100\%
\end{equation}

This measure reflects the reliability of the proposed scheme across diverse operating conditions.

\subsubsection{Spike Economy}

Spike economy measures the efficiency of spike generation. It is defined as the average number of spikes per event across all DERs:
\begin{equation}
\text{Spike Economy} = \frac{S_{\text{total}}}{N_{\text{DERs}}}
\end{equation}

An efficient system should generate minimal spikes during normal load fluctuations but respond with dense, rapid spiking under severe faults. This balance captures both energy savings and sensitivity to disturbances.

\subsection{Simulation Scenarios and Performance Metrics}

System performance was validated across over 300 fault cases and 125 load-change cases using two representative networks: a 3-DER ring and a 4-bus meshed topology (\textit{Supplementary Fig. 2}). Fault types included single-line-to-ground (AG), line-to-line (AB), and three-phase (ABC), with resistances ranging from \( 0.001\,\Omega \) to \( 30\,\Omega \), covering solid short circuits to high-impedance faults. Faults were applied at multiple network points to evaluate spatial selectivity. Load disturbances of \( \pm 40\% \) were applied at DERs and distributed loads to assess immunity against false trips.

In addition to offline simulations, the proposed neuromorphic protection framework was validated in real time on a 3-DER ring microgrid using an OPAL-RT simulator, as detailed in Supplementary Note 9. Each DER is equipped with standard hardware, including local voltage and current sensing and a digital controller executing lightweight LIF neuron updates at microsecond resolution. The disturbance index is computed entirely locally, and spike-based outputs actuate circuit breakers directly, without any communication links.

As shown in Supplementary Fig. 12, load changes do not trigger spiking, whereas faults are detected rapidly—within approximately 10 ms—by the nearest DER, while system frequency remains close to 50 Hz. When breaker actuation is enabled (Supplementary Fig. 13), the earliest spike activates FTTS logic, leading to fast isolation of the faulted line and rapid post-fault frequency stabilization. These real-time results closely mirror the offline simulations and confirm fast, spatially selective fault isolation with negligible sensitivity to normal operating disturbances.

Performance was quantified using three metrics: (i) \textit{tripping latency}, the time from fault onset to breaker operation; (ii) \textit{detection accuracy}, the percentage of cases where the nearest DER cleared the fault within \( 40\,\text{ms} \); and (iii) \textit{spatial selectivity}, whether spikes originated primarily from the DER closest to the fault. Automated scripts extracted these measures and generated qualitative visualization plots (Figs.~3--4), confirming rapid, accurate, and spatially selective protection while maintaining immunity to load changes---demonstrating suitability for real-time, communication-free microgrid protection.

\subsection{Robustness to Parameter Variations and System Scaling}

We tested robustness under variations in neuron model parameters, including spike gain ($k$), membrane time constant ($\tau_m = R_m C_m$), and threshold sensitivity ($\eta$). Across all cases, the system preserved high detection accuracy ($>$98\%) and showed negligible false alarms, even under extreme load changes and diverse fault scenarios. Scalability was further demonstrated by applying the method to different network topologies and DER counts. As illustrated in \textit{Supplementary Fig. 4--6}, the \textit{First-to-Spike} (FTTS) rule consistently identified the DER closest to the fault, ensuring both directional accuracy and spatial selectivity independent of network size or configuration.

A key advantage of this neuromorphic approach is its \textit{communication-free, plug-and-play nature}. No synchronization or central coordination is required, enabling seamless integration of new DERs or topology changes. This decentralization makes the method particularly attractive for future inverter-based microgrids, where protection must remain adaptive and scalable.

\nocite{your-supplementary-citations}

\printbibliography[title={References}]

\section*{Data and Code Availability}

All data generated during this study are provided in the accompanying \textit{Source Data} file. The code, together with the required datasets and detailed instructions to reproduce the results and generate the figures, has been deposited in GitHub: \url{git clone https://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@github.com/saurabhprabhakar/Neuromorphic_Protection.git}, and will be made publicly available upon publication of this paper.

\newpage

\section*{Supplementary Information}

\setcounter{figure}{0} 
\renewcommand{\thefigure}{Supplementary Fig.~\arabic{figure}}

\begin{refsection}

\begin{table}[ht]
\caption*{\textbf{Supplementary Table 1:} Electrical parameters for the ring topology microgrid. All DERs have identical ratings.}
\label{tab:ring_microgrid}
\begin{tabular}{@{}llll@{}}
\toprule
\textbf{Parameter} & \textbf{Symbol} & \textbf{Value} & \textbf{Units} \\
\midrule
Rated Power & $P_{\text{rated}}$ & 10 & kW \\
Line Voltage & $V_L$ & 415 & V (L–L) \\
Frequency & $f$ & 50 & Hz \\
Impedance DER1–DER2 & $R_{12}+jX_{12}$ & $0.7 + j1.884$ & $\Omega$ \\
Impedance DER1–DER3 & $R_{13}+jX_{13}$ & $0.4 + j6.154$ & $\Omega$ \\
Impedance DER2–DER3 & $R_{23}+jX_{23}$ & $1.4 + j3.14$ & $\Omega$ \\
Filter Inductance & $L_f$ & 4 & mH \\
Filter Capacitance & $C_f$ & 200 & $\mu$F \\
DC Link Voltage & $V_{\text{dc}}$ & 1000 & V \\
\botrule
\end{tabular}
\end{table}

\begin{table}[ht]
\caption*{\textbf{Supplementary Table 2:} Electrical parameters for the meshed microgrid topology. All DERs have identical ratings.}
\label{tab:meshed_microgrid}
\begin{tabular}{@{}llll@{}}
\toprule
\textbf{Parameter} & \textbf{Symbol} & \textbf{Value} & \textbf{Units} \\
\midrule
Rated Power & $P_{\text{rated}}$ & 10 & kW \\
Line Voltage & $V_L$ & 415 & V (L–L) \\
Frequency & $f$ & 50 & Hz \\
Impedance DER1–DER2 & $R_{12}+jX_{12}$ & $0.3 + j1.884$ & $\Omega$ \\
Impedance DER1–DER3 & $R_{13}+jX_{13}$ & $0.2 + j6.154$ & $\Omega$ \\
Impedance DER2–DER3 & $R_{23}+jX_{23}$ & $0.7 + j3.14$ & $\Omega$ \\
Impedance DER1–DER4 & $R_{14}+jX_{14}$ & $0.1 + j6.154$ & $\Omega$ \\
Filter Inductance & $L_f$ & 4 & mH \\
Filter Capacitance & $C_f$ & 200 & $\mu$F \\
DC Link Voltage & $V_{\text{dc}}$ & 1000 & V \\
\botrule
\end{tabular}
\end{table}

\newpage
\begin{table*}[ht]
\caption*{\textbf{Supplementary Table 3:} Neuromorphic LIF neuron parameters used to model DER units. Variations across neurons are introduced to study parameter sensitivity.}
\label{tab:lif_neuron_params}
\begin{tabular}{@{}lllll@{}}
\toprule
\textbf{Parameter} & \textbf{Symbol} & \textbf{Neuron 1} & \textbf{Neuron 2} & \textbf{Neuron 3} \\
\midrule
Membrane Capacitance & $C_m$ & 200 $\mu$F & 250 $\mu$F & 100 $\mu$F \\
Membrane Resistance & $R_m$ & 0.432 $\Omega$ & 0.644 $\Omega$ & 0.470 $\Omega$ \\
Membrane Time Constant & $\tau = R_m C_m$ & 86.4 $\mu$s & 161 $\mu$s & 47 $\mu$s \\
Base Threshold Voltage & $U_0$ & 17.9 V & 17.9 V & 17.9 V \\
Threshold Decay Factor & $\eta$ & 0.0156 & 0.0081 & 0.0306 \\
Threshold Damping & $\lambda$ & 0.6 & 0.6 & 0.6 \\
Spike Sampling Interval & $\Delta t$ & 3.125 $\mu$s & 3.125 $\mu$s & 3.125 $\mu$s \\
Disturbance Weight (Voltage) & $\alpha$ & 1.0 & 1.0 & 1.0 \\
Disturbance Weight (Current) & $\beta$ & 0.5 & 0.5 & 0.5 \\
Disturbance Weight (Power rate) & $\gamma$ & 0.005 & 0.005 & 0.005 \\
Spike Gain Factor & $k$ & 20 & 20 & 20 \\
\botrule
\end{tabular}
\end{table*}

\begin{table*}[ht]
\centering
\caption*{\textbf{Supplementary Table 4.} Comparison of the proposed neuromorphic framework with representative microgrid protection schemes.}
\label{tab:protection_comparison}
\small
\renewcommand{\arraystretch}{1.25}
\setlength{\tabcolsep}{4pt}

\begin{tabular}{p{0.6cm} p{2.5 cm} p{2.5 cm} p{1.3 cm} p{1.3 cm} p{3.0cm}}
\toprule
\rowcolor{gray!15}
& \textbf{Protection architecture and representative scheme} 
& \textbf{Decision logic} 
& \textbf{Comm. req.}  
& \textbf{Detection speed}    
& \textbf{Key limitations} \\
\midrule

\rowcolor{gray!8}
\multicolumn{6}{l}{\textbf{1. Conventional decentralized schemes}} \\[0.5ex]

1a & Conventional: Directional overcurrent (DOCR) 
   & Fixed or adaptive current thresholds 
   & No 
   & $>50$ ms 
   & Poor selectivity and coordination in meshed or inverter-dominated grids \\[0.7ex]

1b & Conventional: Adaptive harmonic relay 
   & Harmonic content / THD-based indices 
   & No 
   & 20--40 ms 
   & Sensitive to non-linear loads and inverter switching noise; may mis-operate under benign harmonics \\[1.2ex]

\rowcolor{gray!8}
\multicolumn{6}{l}{\textbf{2. Communication-based hybrid schemes}} \\[0.5ex]

2a & Communication-based: PMU-based consensus / wide-area schemes 
   & Phasor state estimation and consensus rules 
   & Yes (GPS sync.) 
   & $<20$ ms 
   & Strong dependence on reliable communication and time-synchronization; exposed to latency, packet loss and cyber attacks \\[0.7ex]

2b & Communication-based: Differential protection (87L) 
   & Current differential comparison between line ends 
   & Yes (fiber / 5G) 
   & $<10$ ms 
   & High installation cost; protection is lost if the communication channel fails \\[1.2ex]

\rowcolor{gray!8}
\multicolumn{6}{l}{\textbf{3. Emerging decentralized schemes}} \\[0.5ex]

3a & Emerging decentralized: Traveling-wave (TW) protection 
   & Wavefront arrival time and correlation 
   & No (local only) 
   & $<5$ ms 
   & Requires MHz-range sampling and high-end hardware; sensitive to noise and modelling errors \\[1.2ex]

\rowcolor{blue!8}
\multicolumn{6}{l}{\textbf{4. Proposed neuromorphic protection}} \\[0.5ex]

4a & \textbf{Proposed: Neuromorphic FTTS} 
   & \textbf{Spike latency and adaptive threshold encoding of disturbance index} 
   & \textbf{No} 
   & \textbf{$<15$ ms (sub-cycle)} 
   & \textbf{Requires tuning of neuron parameters ($\tau$, $k$, threshold gains)} \\

\bottomrule
\end{tabular}
\end{table*}

\clearpage

\begin{figure*} 
    \centering
    \includegraphics[scale=0.65]{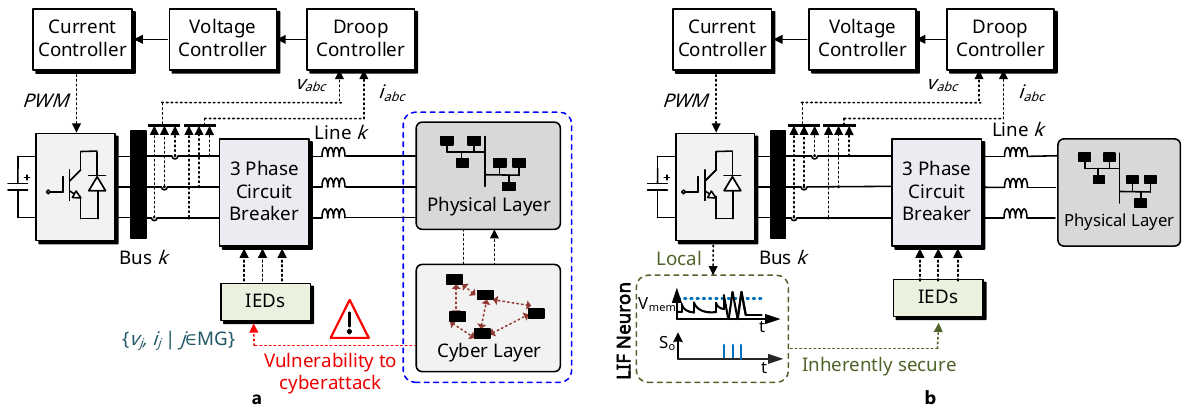}
    \caption*{\textbf{Supplementary Fig.1: Moving beyond traditional hierarchical cyber-physical microgrids.} 
This figure compares conventional and neuromorphic approaches to microgrid protection. 
In (a), traditional Intelligent Electronic Device (IED)-based breakers depend on remote data from a central controller or communication layer to operate, making them vulnerable to delays, cyber-attacks, or synchronization errors. 
In contrast, (b) shows the proposed co-transfer architecture, where each Distributed Energy Resource (DER) is modeled as a Leaky Integrate-and-Fire (LIF) neuron. 
These neurons monitor local disturbances and trigger protection actions independently, removing the need for centralized communication and enabling faster, decentralized fault isolation.}

    \label{fig:control_schematic}
    
\end{figure*}

\begin{figure*}
    \centering
    \includegraphics[scale=0.65]{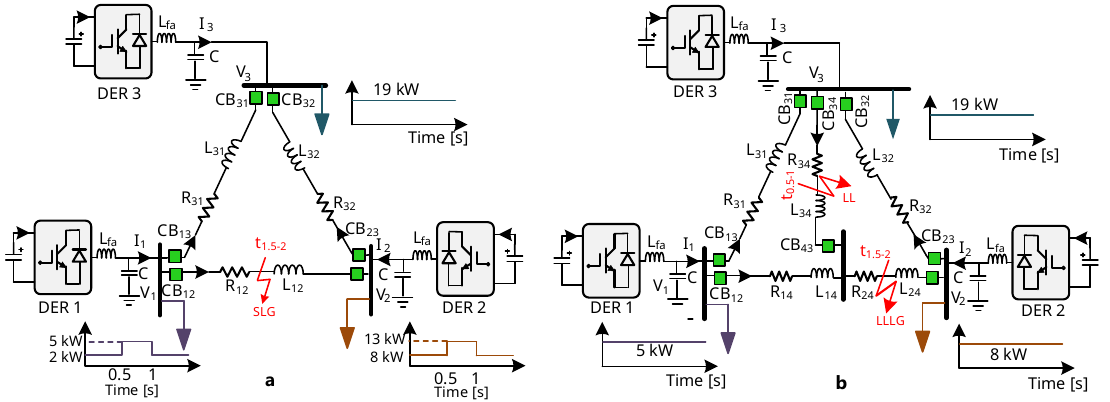}
    \caption*{\textbf{Supplementary Fig. 2: Microgrid topologies used for validating the neuromorphic protection framework.} 
Two representative configurations are modeled to evaluate system performance under varying network structures. 
In (a), Case I shows a three-bus, three-DER AC microgrid arranged in a ring topology, enabling bidirectional fault propagation paths. 
In (b), Case II extends this setup to a four-bus, three-DER system with an additional tie-line, forming a meshed topology. 
This configuration introduces redundant paths and higher connectivity, allowing analysis of fault discrimination and selectivity in more complex grid conditions.}

    \label{fig:control_schematic}
    \vspace{-2mm}
\end{figure*}
\begin{figure*}
    \centering
    \includegraphics[scale=0.85]{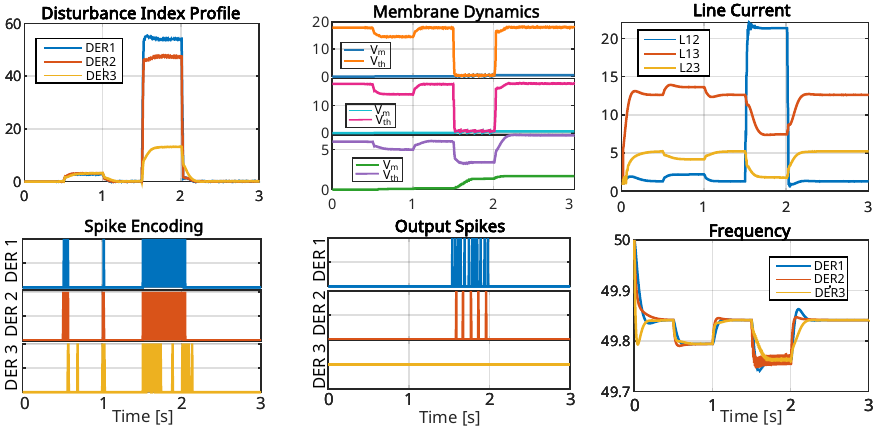}
    
    \caption*{\textbf{Supplementary Fig. 3: Spike-based encoding response in a 3-bus ring microgrid without protection.} 
\textbf{Disturbance Profile:} The system experiences a moderate load increase followed by a single-line-to-ground (SLG) fault, resulting in a sharp rise in the disturbance index $D_i(t)$. 
\textbf{Spike Encoding:} During load changes, the adaptive threshold \( V_{\text{th}}(t) \) adjusts dynamically, modulating in response to the disturbance and suppressing unnecessary spikes.
When the fault occurs, spike intervals shorten and neurons begin firing rapidly. 
\textbf{Membrane Dynamics:} The membrane voltage $V_\text{m}(t)$ climbs more quickly during the fault and crosses the threshold, indicating a strong neuronal response. 
\textbf{Output Spikes:} Spikes occur immediately after the fault onset, reflecting early detection by the neuromorphic system. 
\textbf{Line Current:} With protection disabled, the fault causes abnormal current to propagate across all lines. 
\textbf{Frequency:} The fault leads to frequency deviations across the network.This figure demonstrates that while the system detects faults promptly through spiking activity, the absence of protection allows the disturbance to spread—underscoring the necessity of fast and local isolation mechanisms like First-To-Spike (FTTS) logic.}
    \label{fig:control_schematic}
    \vspace{-2mm}
\end{figure*}

\begin{figure*}
    \centering
    \includegraphics[scale=0.9]{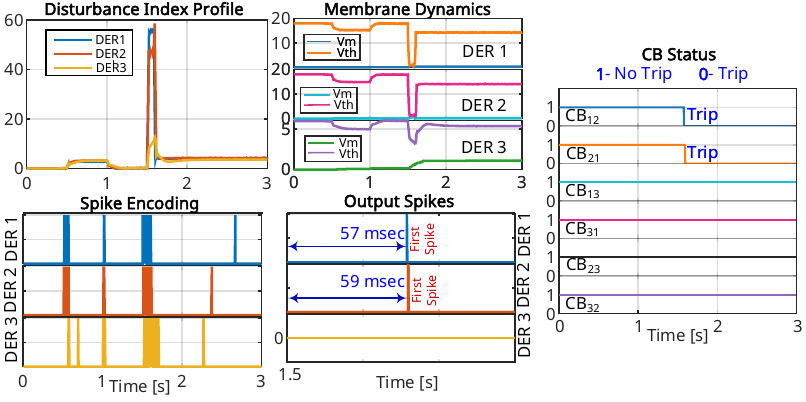}
    
    \caption*{\textbf{Supplementary Fig. 4: How spike-based coordination with FTTS enables fast and directional fault isolation.} 
\textbf{Disturbance Profile:} A fault is introduced in the same 3-bus ring microgrid, but this time FTTS logic is active. 
\textbf{Spike Encoding:} The DER that is closest to the fault detects the disturbance first and emits an early spike, clearly ahead of the full development of fault current. 
\textbf{Membrane Dynamics:} Only the membrane voltage $V_\text{m}(t)$ of the impacted DER crosses its adaptive threshold $V_{\text{th}}(t)$, while the other DERs remain below threshold—showing the system’s built-in spatial selectivity. 
\textbf{Output Spikes:} The first spike from the most affected DER triggers Circuit Breakers 12 and 21, isolating the faulted line from both sides within just 1.5 cycles. 
\textbf{Line Current:} After isolation, fault current is suppressed quickly and does not spread to the rest of the network. 
\textbf{Frequency:} The frequency in unaffected areas remains stable, confirming that the fault has been contained.
}
    \label{fig:control_schematic}
    \vspace{-2mm}
\end{figure*}

\begin{figure*}
    \centering
    \includegraphics[scale=0.9]{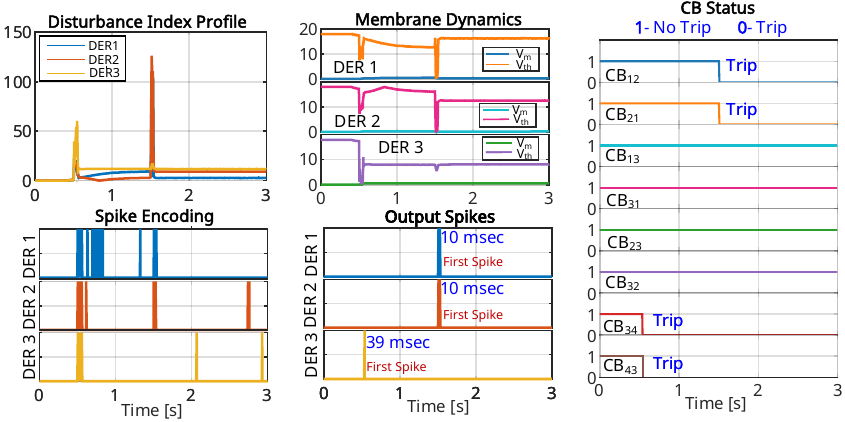}
    
    \caption*{\textbf{Supplementary Fig. 5: Neuromorphic protection adapts to multiple faults in a meshed microgrid using localized spike-based responses.} 
\textbf{Disturbance Profile:} Two sequential faults—a line-to-line (LL) fault followed by a more severe three-phase-to-ground (LLLG) fault—are introduced in a 4-bus meshed topology. 
\textbf{Spike Encoding:} Each fault generates a distinct spike response: the LLLG fault causes faster and denser spikes due to a higher disturbance index $D(t)$, compared to the LL fault. 
\textbf{Membrane Dynamics:} The membrane potentials rise sharply and cross the adaptive threshold more quickly for the second fault, demonstrating severity-aware encoding. 
\textbf{Output Spikes:} Spikes from the nearest DERs initiate fast, localized action. 
\textbf{Circuit Breaker Status:} For the first fault, CB34 and CB43 trip to isolate the affected line. When the second fault occurs on a different line, CB12 and CB21 trip accordingly—ensuring that only faulted segments are disconnected while the rest of the grid remains operational.}
    \label{fig:control_schematic}
    \vspace{-2mm}
\end{figure*}

\begin{figure*}
    \centering
    \includegraphics[scale=0.95]{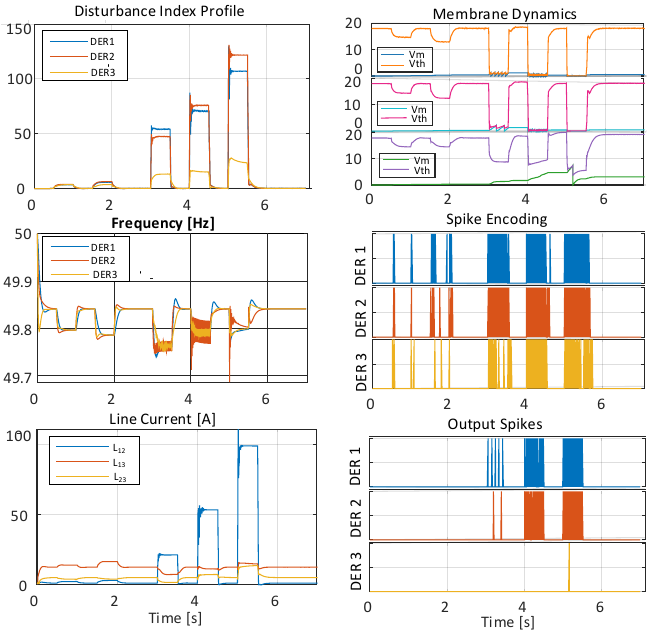}
    
    \caption*{\textbf{Supplementary Fig. 6: Simple and clear view of how spike-based protection adapts to fault severity and location across all DERs.} 
\textbf{Disturbance Index:} In three fault cases—AG, ABG, and ABCG—the disturbance index $D(t)$ rises sharply only during faults, not during normal conditions. 
\textbf{Spike Encoding:} This rise leads to dense and early input spikes from the DER closest to the fault (b). 
\textbf{Membrane Dynamics:} The membrane potential $V_\text{m}(t)$ increases and crosses the adaptive threshold $V_{\text{th}}(t)$ only under serious fault conditions (c). 
\textbf{Output Spikes:} As a result, spikes are triggered quickly—at 58 ms for AG and just 10 ms for the more severe ABCG fault (d), showing how spike timing depends on fault intensity. 
\textbf{Line Current and Frequency:} Current and frequency signals (e, f) show the disturbance clearly at faulted DERs, while unaffected DERs stay silent—no unnecessary spikes. This demonstrates that the First-To-Spike (FTTS) logic helps detect faults quickly, accurately, and without needing any communication. The system responds only when needed, saving energy and ensuring correct circuit breaker operation based on local information.}
    \label{fig:control_schematic}
    \vspace{-2mm}
\end{figure*}

\begin{figure*}
    \centering
    \includegraphics[scale=0.80]{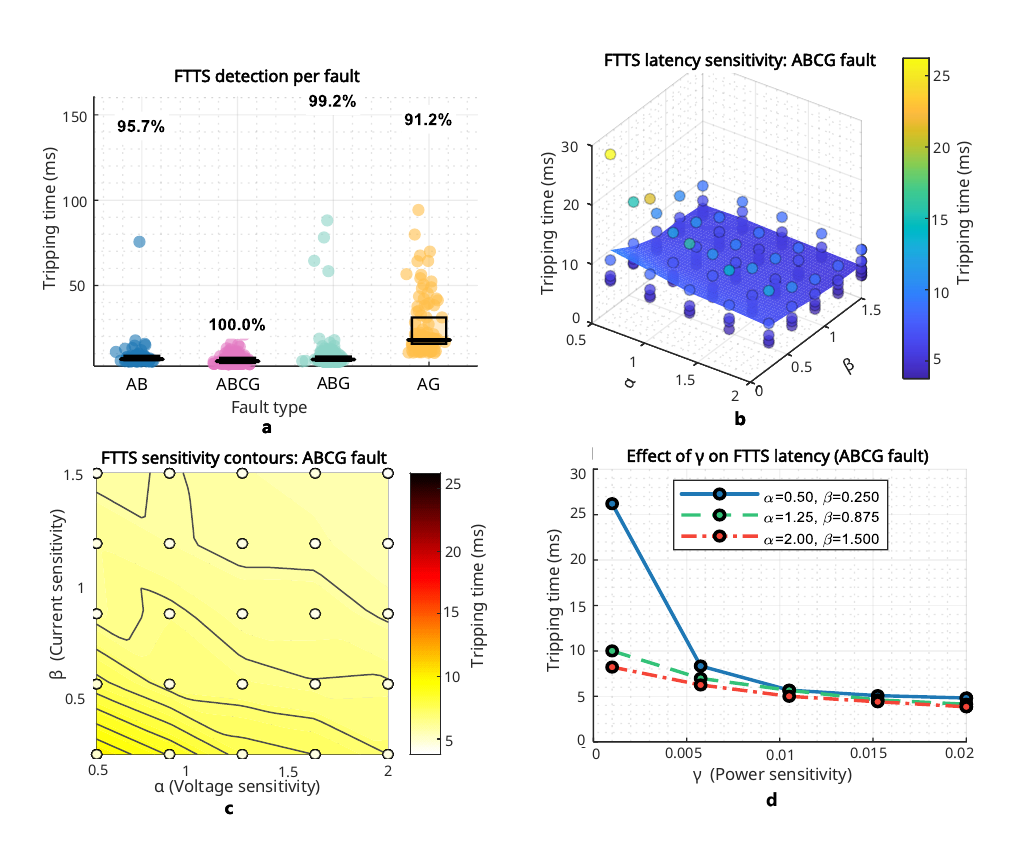}
    
    \caption*{\textbf{Supplementary Fig. 7: Sensitivity of the neuromorphic FTTS protection to disturbance-index weights.} 
\textbf{(a)} Scatter–box (violin) distributions of FTTS tripping time for different fault types, with overlaid detection accuracy, demonstrate reliable fault detection across all classes while maintaining fast, millisecond-scale latency. 
\textbf{(b)} Interpolated 3D response surface of tripping time versus voltage- and current-sensitivity weights ($(\alpha, \beta)$) for ABCG faults shows a smooth, well-behaved dependence and consistently low latency around the nominal operating point. 
\textbf{(c)} Corresponding $(\alpha, \beta)$ heatmap highlights a broad plateau of admissible parameter combinations, with worst-case tripping times remaining well below half a cycle at 50 Hz. 
\textbf{(d)} Tripping-time variation with power-sensitivity weight $\gamma$ for representative $(\alpha, \beta)$ settings indicates only weak dependence on $\gamma$, with latency staying below $\sim$25 ms across the full sweep, underscoring the robustness of the proposed FTTS scheme to coefficient variations.}
    \label{fig:control_schematic}
    \vspace{-2mm}
\end{figure*}

\begin{figure*}
    \centering
    \includegraphics[scale=0.65]{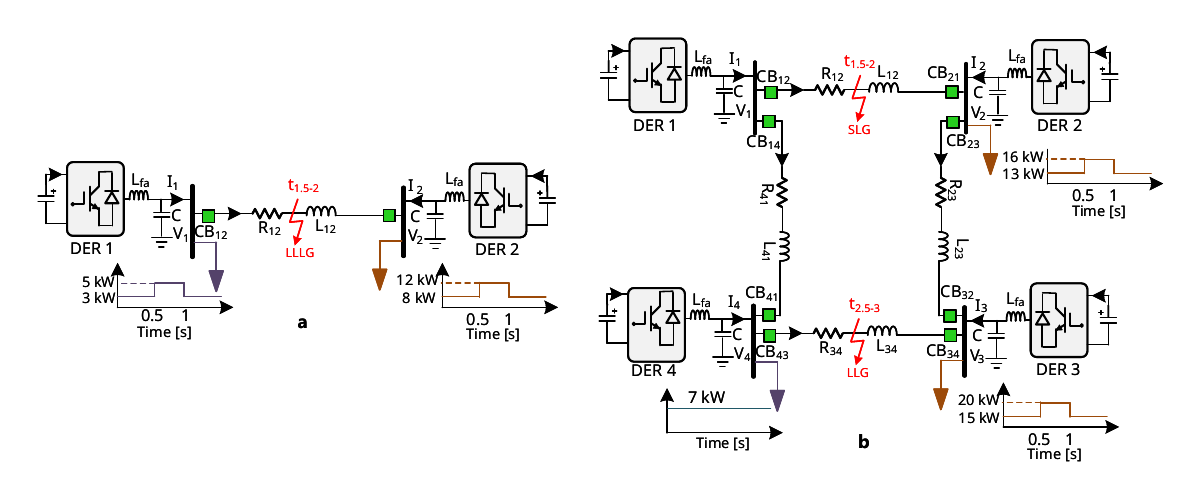}
    
    \caption*{
    \textbf{Supplementary Fig. 8:} 
    \textbf{a) 3-bus, 3-DER ring system:} Configuration used to test fault isolation and fault current suppression in a more densely connected network, with increased redundancy over the 2-bus system.\textbf{b) 4-bus, 4-DER ring microgrid:} Used to evaluate scalability and fault detection accuracy in a more complex system with four buses and four distributed energy resources (DERs).
}

    \label{fig:control_schematic}
    \vspace{-2mm}
\end{figure*}

\begin{figure*}
    \centering
    \includegraphics[scale=0.80]{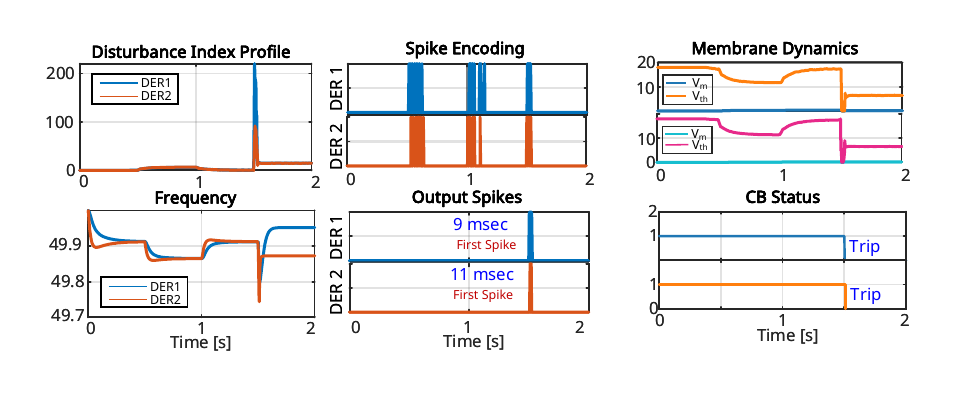}
    
   \caption*{
    \textbf{Supplementary Fig. 9: Neuromorphic FTTS protection response in a 2-bus radial microgrid.} 
    \textbf{Disturbance sequence:} A load change is applied from 0.5–1.0 s, followed by a three-phase-to-ground (LLLG) fault at 1.5 s. \textbf{Spike dynamics:} During the load change, the disturbance index increases moderately but remains below the adaptive threshold, preventing spurious spiking. \textbf{Fault response:} Upon fault inception, the membrane potential rapidly crosses its threshold, producing an early spike that triggers correct breaker action. \textbf{System behavior:} Fault current is quickly suppressed, confirming correct and selective tripping even in the simplest radial topology.
}

    \label{fig:control_schematic}
    \vspace{-2mm}
\end{figure*}

\begin{figure*}
    \centering
    \includegraphics[scale=0.80]{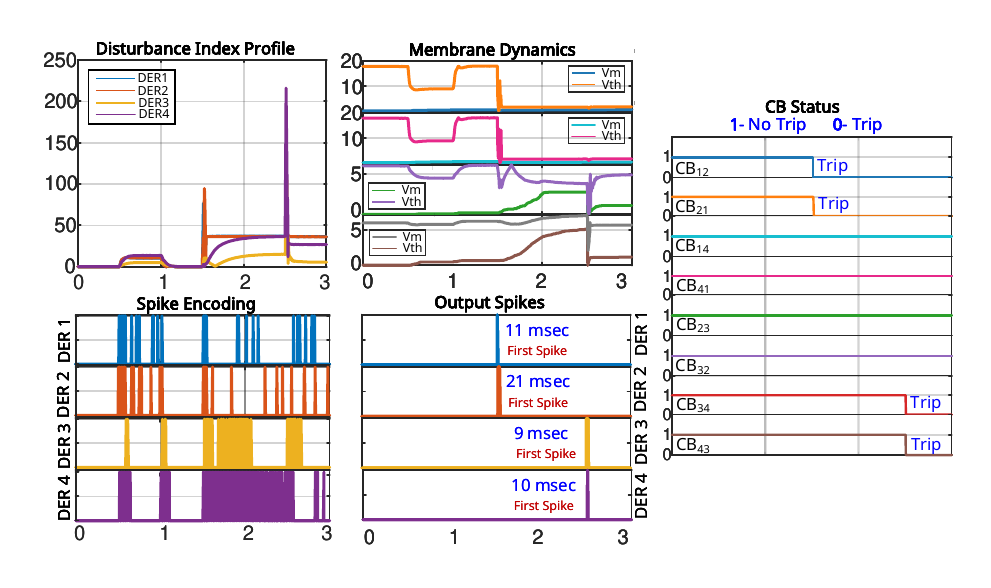}
    
    \caption*{
    \textbf{Supplementary Fig. 10: Multi-event fault response in a 4-bus, 4-DER ring microgrid.} \textbf{Disturbance sequence:} A load change occurs between 0.5–1.0 s, an SLG fault is introduced from 1.5–2.0 s, and a more severe LLLG fault follows from 2.5–3.0 s. \textbf{Selectivity:} Load disturbances do not trigger protection, while both faults are detected and isolated correctly. \textbf{Severity awareness:} The LLLG fault induces faster and denser spiking compared to the SLG fault, reflecting higher disturbance intensity.  \textbf{Scalability:} Correct fault isolation across multiple DERs confirms that the neuromorphic FTTS framework scales naturally with network size and complexity.
}

    \label{fig:control_schematic}
    \vspace{-2mm}
\end{figure*}

\newpage

\clearpage
\begin{center}
\begin{algorithm}[H]
\footnotesize
\caption{Neuromorphic FTTS Protection for 3-DER Islanded Microgrid}
\label{alg:neuromorphic_ftts}

\begin{algorithmic}[1]

\Require DERs $i \in \{1, 2, 3\}$; sampling period $\Delta t$; tuning parameters $k, \alpha, \beta, \gamma, \eta, \lambda$; noise floor $D_{\min}$

\Ensure Trip time $t_{\text{FTTS}}$ and tripped DER index $i^*$
\vspace{0.5em}
\State \textbf{Initialization:}
\For{each DER $i$}
    \State $V_{0,i} \gets \text{measure\_voltage}(i)$, $I_{0,i} \gets \text{measure\_current}(i)$, $P_{0,i} \gets \text{measure\_power}(i)$
    \State $V_{m,i}(0) \gets 0$ \textit{(membrane potential)}, $n_i \gets 0$ \textit{(spike counter)}
\EndFor
\vspace{0.5em}
\State \textbf{Main Loop:} for each time step $t = \Delta t, 2\Delta t, 3\Delta t, \ldots$ do
\vspace{0.3em}
\State \underline{\textbf{Step 1: Local Measurement}}
\For{$i \in \{1, 2, 3\}$}
    \State $V_i(t) \gets \text{acquire\_voltage}(i)$, $I_i(t) \gets \text{acquire\_current}(i)$, $P_i(t) \gets \text{compute\_power}(V_i(t), I_i(t))$
\EndFor
\vspace{0.3em}
\State \underline{\textbf{Step 2: Disturbance Computation}}
\For{$i \in \{1, 2, 3\}$}
    \State $\Delta V_i(t) \gets |V_i(t) - V_{0,i}|$, $\Delta I_i(t) \gets |I_i(t) - I_{0,i}|$, $\Delta P_i(t) \gets |P_i(t) - P_{0,i}|$
    \State $D_i(t) \gets \alpha \cdot \Delta V_i(t) + \beta \cdot \Delta I_i(t) + \gamma \cdot \Delta P_i(t)$ (Eq. 3.1)
    \If{$D_i(t) < D_{\min}$} \State $D_i(t) \gets 0$ \EndIf
\EndFor
\vspace{0.3em}
\State \underline{\textbf{Step 3: Spike Interval \& Input Encoding}}
\For{$i \in \{1, 2, 3\}$}
    \If{$D_i(t) > 0$}
        \State $T_{s,i}(t) \gets \dfrac{1}{1 + k \cdot D_i(t)}$ (Eq. 3.2)
        \If{$(t - n_i \cdot T_{s,i}(t)) \geq T_{s,i}(t)$}
            \State $S_i(t) \gets 1$ \textit{(input spike)}, $n_i \gets n_i + 1$
        \Else
            \State $S_i(t) \gets 0$
        \EndIf
    \Else
        \State $S_i(t) \gets 0$
    \EndIf
\EndFor
\vspace{0.3em}
\State \underline{\textbf{Step 4: Adaptive Threshold Update}}
\For{$i \in \{1, 2, 3\}$}
    \State $V_{\text{th},i}(t) \gets V_0 \cdot \exp\big(-\eta(\Delta V_i(t) + \Delta I_i(t) + \Delta P_i(t))\big) + \lambda \cdot V_{m,i}(t)$ (Eq. 3.3)
\EndFor
\vspace{0.3em}
\State \underline{\textbf{Step 5: LIF Membrane Dynamics}}
\For{$i \in \{1, 2, 3\}$}
    \State $D_{\text{input},i}(t) \gets S_i(t)$ \textit{(or weighted input from spike train)}
    \State $\dfrac{dV_{m,i}}{dt} \gets \dfrac{-V_{m,i}(t)/R_m + D_{\text{input},i}(t)}{C_m}$ (Eq. 3.4)
    \State $V_{m,i}(t + \Delta t) \gets V_{m,i}(t) + \left(\dfrac{dV_{m,i}}{dt}\right) \cdot \Delta t$ \textit{(Euler integration)}
\EndFor
\vspace{0.3em}
\State \underline{\textbf{Step 6: Output Spike Detection}} $\text{fired\_neurons} \gets \emptyset$
\For{$i \in \{1, 2, 3\}$}
    \If{$V_{m,i}(t + \Delta t) \geq V_{\text{th},i}(t)$ and $i \notin \text{fired\_neurons}$}
        \State $t_i^{(1)} \gets t + \Delta t$ \textit{(first spike time)}, $\text{fired\_neurons} \gets \text{fired\_neurons} \cup \{i\}$, $V_{m,i}(t + \Delta t) \gets 0$
    \EndIf
\EndFor
\vspace{0.3em}
\State \underline{\textbf{Step 7: First-To-Spike (FTTS) Selection}}
\If{$\text{fired\_neurons} \neq \emptyset$}
    \State $i^* \gets \arg\min\limits_{i \in \text{fired\_neurons}} \{t_i^{(1)}\}$, $t_{\text{FTTS}} \gets t_{i^*}^{(1)}$
    \State \underline{\textbf{Step 8: Trip Action \& Isolation}} $\text{Issue\_Trip\_Command}(\text{Breaker}_{i^*})$
    \State \Return $(t_{\text{FTTS}}, i^*)$ \textit{(exit with protection action)}
\EndIf
\vspace{0.3em}
\end{algorithmic}
\end{algorithm}
\end{center}

\clearpage

\newpage

\section*{Supplementary Note 1 — Spike-Latency Encoding via Disturbance Index in LIF-Based Protection}

This section explains the core mathematical foundation of the proposed spike-based neuromorphic protection scheme. In this framework, each distributed energy resource (DER) is modeled as a Leaky Integrate-and-Fire (LIF) neuron. These neuron-like DERs monitor local voltage, current, and power signals, and generate spikes to indicate disturbances. The timing of each spike, also known as spike latency is influenced by a disturbance index, which helps quantify how abnormal the local electrical conditions are.

\paragraph{Disturbance Index Definition} To detect 
disturbances in the microgrid, each DER continuously compares its measured electrical quantities with their steady-state (pre-fault) values, denoted as \( V_0 \), \( I_0 \), and \( P_0 \). The deviations are calculated as:

\setcounter{equation}{0}

\renewcommand{\theequation}{S\arabic{equation}}

\begin{equation}
\Delta V(t) = \left| V(t) - V_0 \right|, \quad
\Delta I(t) = \left| I(t) - I_0 \right|, \quad
\Delta P(t) = \left| P(t) - P_0 \right|
\label{supp_eq:S1}
\end{equation}

These deviations are then combined into a single scalar value called the disturbance index, defined as:
\begin{equation}
D(t) = \alpha \left| \Delta V(t) \right| + \beta \left| \Delta I(t) \right| + \gamma \left| \Delta P(t) \right|
\end{equation}

Here, the constants $\alpha$, $\beta$, and $\gamma$ are tuning parameters between 0 and 1. They control how much each signal contributes to the overall disturbance index. This makes the system flexible, allowing it to focus more on voltage, current, or power deviations depending on the setup. The disturbance index simplifies the complex multi-signal input into one single quantity that can drive spike-based decisions efficiently.

\paragraph{Spike Latency Modulation} The disturbance index is then used to control how often a neuron (DER) spikes. This is done using an inverse-time encoding rule, where the time between spikes, known as the spike interval, is given by:
\begin{equation}
T_{s}(t) = \frac{1}{1 + k D(t)}
\end{equation}

The parameter $k$ is a positive gain that determines how sensitive the system is to disturbances. A higher value of $D(t)$ results in a smaller $T_{s}(t)$, meaning spikes are generated more frequently. In simple terms, the more severe the disturbance, the faster the neuron fires. This mirrors how traditional inverse-time relays behave: bigger faults cause faster trip actions. Here, that same behavior is achieved using biologically inspired spike timing.

\paragraph{Spike Train Generation} Once the spike interval is known, a spike train is created to represent when spikes occur. This is implemented using a step function known as the Heaviside function, which switches ON when a threshold is crossed. The spike train is written as:
\begin{equation}
S_i(t) = \sum_{n} \Theta\left(t - n T_{s,i}(t)\right) \quad \text{where}, \quad \Theta(x) =
\begin{cases}
1, & x \geq 0 \\
0, & x < 0
\end{cases}
\end{equation}

where the subscript \( i \) denotes the \( i \)-th Distributed Energy Resource (DER), \( T_{s,i}(t) \) is the spike interval at that DER, and \( \Theta(\cdot) \) is the Heaviside step function generating the spike train \( S_i(t) \).This mathematical expression simply says that a spike occurs every $T_{s,i}(t)$ seconds. The spike train behaves like an event-driven signal: quiet during normal conditions, and active when disturbances are present. Because it is binary and time-based, this type of spike signal is easy to process in real-time digital or analog hardware.

\paragraph{Summary} In summary, this spike-encoding method converts real-world voltage, current, and power deviations into biologically inspired spike patterns. These spikes can be used to detect faults quickly and locally, without needing any communication between DERs. This forms the foundation of our First-to-Spike (FTTS) protection logic, which enables fast, robust, and distributed fault detection.

\newpage
\subsection*{Supplementary Note 2 — Adaptive Threshold Modulation for Selective Spike Generation}

To improve the reliability of the neuromorphic protection system, each Leaky Integrate-and-Fire (LIF) neuron uses an adaptive threshold. This means the neuron will only spike (i.e., react) when the electrical disturbance is strong enough to be considered meaningful. This approach helps the system avoid false alarms during small, routine load changes, while still reacting quickly to actual faults. In real neurons, the firing threshold can change based on the intensity of input signals. Inspired by this, our system allows the spike threshold to adjust in real-time depending on how much the local voltage, current, or power changes. When a small disturbance occurs, the threshold stays high so that no spike is generated. But when a significant fault happens, the threshold quickly lowers, making it easier for the membrane voltage to cross it and trigger a spike. The adaptive threshold is calculated using the following formula:
\begin{equation}
V_{\text{th}}(t) = V_0 \cdot e^{-\eta(\Delta V(t) + \Delta I(t) + \Delta P(t))} + \lambda V_{\text{m}}(t)
\end{equation}

Here, $V_0$ is the default threshold voltage. The first part of the formula reduces the threshold when the disturbance increases, and the second part adds memory by considering the current value of the membrane potential $V(t)$. The parameter $\eta$ controls how sensitive the threshold is to changes in voltage, current, and power, while $\lambda$ helps prevent unnecessary spikes by dampening the response.

This dynamic threshold works alongside the membrane potential, which evolves over time according to the disturbance:
\begin{equation}
C_m \frac{dV_{\text{m}}(t)}{dt} = -\frac{V_{\text{m}}(t)}{R_m} + D_{\text{input}}(t)
\end{equation}
A spike is generated when the membrane potential $V_{\text{m}}(t)$ becomes greater than or equal to the threshold $V_{\text{th}}(t)$:
\begin{equation}
V_{\text{m}}(t) \geq V_{\text{th}}(t)
\end{equation}

This method allows the neuron to spike only during actual, serious disturbances and stay quiet during normal operation. As a result, the system becomes highly selective, energy-efficient, and robust—reacting only when needed, just like a biological neuron filters out background noise but responds to real danger.

\newpage

\subsection*{Supplementary Note 3 — First-To-Spike Logic and Decentralized Fault Localization}

This section explains the First-To-Spike (FTTS) logic, which allows fast, fully decentralized fault detection in the neuromorphic protection framework. Inspired by how biological neurons respond to stimuli, FTTS identifies the DER that spikes first when a disturbance occurs—this DER is considered closest to the fault and triggers the associated circuit breaker.

In nature, neurons that receive the strongest input fire first. Similarly, in our system, the DER experiencing the highest local disturbance emits the earliest spike. This method eliminates the need for communication, synchronization, or global coordination. Mathematically, suppose each DER neuron $i$ produces spikes at times $t_i(n)$, where $n$ is the spike number. The first spike among all DERs is:

\begin{equation}
t_{\text{FTTS}} = \min_i \{ t_i(1) \}
\end{equation}

The DER with this first spike (denoted $i^*$) immediately sends a trip signal to its breaker:

\begin{equation}
\text{Breaker}_{i^*} \leftarrow \text{Trip at } t_{i^*}(1)
\end{equation}

This simple rule ensures only the nearest DER to the fault acts, keeping the response local, fast, and selective—just like in a biological nervous system.

\newpage
\subsection*{Supplementary Note 4 — Multi-line Disturbance Aggregation and Spatial Selectivity}

In AC microgrids, especially those with ring or meshed topologies, a single DER is often connected to multiple lines. This presents a key challenge: \textit{how can one DER neuron figure out which specific line is experiencing a fault, based only on local information?}

To solve this, we introduce a method called \textit{multi-line disturbance aggregation}, which allows each DER to monitor all its connected lines separately and still make a spatially selective protection decision. The idea is that each line connected to a DER contributes its own disturbance information, and the DER processes them in parallel to determine which line is most affected.

When a DER (denoted as node $i$) is connected to $N$ lines, For each connected line $j$, it computes a disturbance index $D_{i,j}(t)$, which is then used to calculate the spike interval:

\begin{equation}
T_{s,i,j}(t) = \frac{1}{1 + k D_{i,j}(t)}
\end{equation}

This spike interval defines how often spikes occur from that line’s input. Each of these is then converted into a spike train:

\begin{equation}
S_{i,j}(t) = \sum_{n} \Theta \left( t - n T_{s,i,j} \right)
\end{equation}

The total spike activity observed by the DER is the sum of all the spike trains from its connected lines:

\begin{equation}
S^{\text{DER}}_i(t) = \sum_{j=1}^{N} S_{i,j}(t)
\end{equation}

This aggregated signal is used by the LIF neuron's membrane, which integrates the combined disturbance as it evolves over time:

\begin{equation}
C_m \frac{dV_{\text{m}}(t)}{dt} = -\frac{V(t)}{R_m} + \sum_{j=1}^{N} D_{i,j}(t)
\end{equation}

In this setup, the line with the strongest disturbance causes more frequent spikes, raising the membrane voltage faster and crossing the spike threshold sooner. This naturally leads the neuron to prioritize the most affected line without any external direction logic.To formally identify the most disturbed line, we define a line priority index $\Pi_{i,j}$ for each line $j$ as:

\begin{equation}
\Pi_{i,j} = \frac{1}{T_{s,i,j}} = 1 + k D_{i,j}(t)
\end{equation}

The line with the highest priority index is the most likely fault location:

\begin{equation}
j^* = \arg\max_j \Pi_{i,j}
\end{equation}

This mechanism enables the system to select the faulted line purely based on spike timing. It effectively replaces traditional directional relays with a spike-based decision process, making the protection logic simpler, faster, and communication-free.

\newpage
\subsection*{Supplementary Note 5 — Adaptive Thresholding for False Trip Suppression and Fault Sensitivity}

In the proposed spike-based protection system, each DER uses a dynamic threshold to decide when to generate a spike. Unlike fixed-threshold systems, this adaptive threshold changes based on the current electrical conditions. The main goal is to avoid false trips during normal load changes, while still reacting quickly to actual faults.

The threshold, denoted as $U_{\text{th}}(t)$, depends on how much the voltage, current, and power deviate from their normal values. When these values stay close to nominal—as in the case of small load fluctuations—the threshold remains high, preventing the neuron from spiking unnecessarily. However, when a large disturbance occurs, such as a fault, the threshold rapidly decreases, enabling the membrane potential to cross it and generate a spike.

Mathematically, the adaptive threshold is given by:

\begin{equation}
V_{\text{th}}(t) = V_0 \cdot e^{-\eta \left( \Delta V(t) + \Delta I(t) + \Delta P(t) \right)} + \lambda V_{\text{m}}(t)
\end{equation}

Here:
\begin{itemize}
  \item $V_0$ is the base threshold level under normal operating conditions,
  \item $\eta$ controls the sensitivity of the threshold to electrical disturbances,
  \item $\lambda$ introduces a memory component by linking the threshold to the current membrane voltage $V_{\text{m}}(t)$.
\end{itemize}

This formulation ensures that minor load variations do not cause false spikes, as the threshold remains elevated. However, during real faults, significant deviations in voltage, current, or power reduce the threshold, making it easier for $V_{\text{m}}(t)$ to reach it and generate a spike.

Once a spike is triggered, the connected circuit breaker responds, isolating the fault quickly and locally.

Overall, the adaptive threshold mechanism enhances the reliability and selectivity of the neuromorphic protection system. It ensures that the neuron remains silent during benign conditions but reacts decisively to critical disturbances, mimicking the behavior of biological neurons that filter out background noise while responding to meaningful stimuli.

\newpage
\subsection*{Supplementary Note 6 — Scaling Behavior of Spike Latency via the Inverse Disturbance Law}

In the neuromorphic protection framework, each DER uses spike timing to signal how severe a local disturbance is. This behavior follows an inverse relationship: the stronger the disturbance, the shorter the time before a spike is fired. This is controlled by a scaling factor $k$, which acts like a sensitivity knob—adjusting how fast the system reacts to faults.

The inter-spike interval is defined by the following equation:

\begin{equation}
T_{s,i}(t) = \frac{1}{1 + k D_i(t)}
\end{equation}

Here, $D_i(t)$ is the disturbance index (a combination of voltage, current, and power deviations), and $k$ is a gain parameter. When the disturbance is small, the spike comes late. When the disturbance is large, the spike happens quickly. This behavior is analogous to traditional overcurrent relays, where trip time decreases as fault current increases. In inverse-time protection schemes (such as those defined by IEC 60255), trip delay depends on how much the current exceeds a preset threshold. Although the mathematical formulations differ, the core concept remains the same: stronger events cause faster action.

The value of $k$ plays a significant role. A low $k$ results in slower spikes, making the system more selective and less likely to respond to minor disturbances. Conversely, a high $k$ speeds up the response—even for small changes—improving reaction time but increasing the chance of false positives. In practice, choosing $k$ between 10 and 100 provides a good balance between speed and selectivity. For example:
\begin{itemize}
    \item Under normal load variation ($D \sim 10$), the spike delay is $\geq 2$ ms.
    \item During faults ($D \sim 100$), the spike delay reduces to $\leq 0.5$ ms—fast enough for protection.
\end{itemize}

The sensitivity of the spike interval with respect to the disturbance index can be analyzed by differentiating $T_{s,i}(t)$ with respect to $D_i$:

\[
\frac{dT_{s,i}}{dD_i} = \frac{-k}{(1 + k D_i)^2}
\]

This expression shows that as disturbances grow larger, the spike interval becomes less sensitive to small changes. Most of the differentiation—and therefore most of the decision-making—occurs at lower disturbance levels, helping the system detect the onset of faults quickly.

\medskip

\noindent \textbf{In summary}, the inverse relationship between disturbance and spike timing allows the neuromorphic system to mimic traditional relay behavior in a more flexible and event-driven way. The gain parameter $k$ serves as a tuning dial that controls how rapidly the system responds and how effectively it distinguishes between benign and severe conditions. This timing-based protection also supports decentralized fault localization, since the first spike indicates proximity to the fault.

\newpage
\subsection*{Supplementary Note 7 — Analytical Mapping to Inverse-Time Relay Characteristics}

Traditional inverse-time overcurrent relays are widely used in power systems because they trip faster under severe faults and slower under mild disturbances. This characteristic is captured by the following standard equation:

\begin{equation}
t_{\text{trip}} = A \left( \frac{I}{I_{\text{pickup}}} \right)^p - 1
\end{equation}

Here, $I$ is the fault current, $I_{\text{pickup}}$ is the minimum current required to initiate a trip, and $A$, $p$ are tuning parameters that shape the curve.

\vspace{0.5em}
In the proposed spike-based neuromorphic framework, a similar inverse-time behavior naturally emerges. The spiking delay of a DER neuron is inversely related to its disturbance index $D_i(t)$, which combines deviations in voltage, current, and power. The spike interval is given by:

\begin{equation}
T_{s,i}(t) = \frac{1}{1 + k D_i(t)} = \frac{1}{1 + k(\alpha |\Delta V(t)| + \beta |\Delta I(t)| + \gamma |\Delta P(t)|)}
\end{equation}

When a fault occurs, the DER neuron generates spikes as soon as its membrane potential $U(t)$ crosses the adaptive threshold $U_{\text{th}}(t)$. The effective tripping time can be approximated as a small multiple of the spike interval:

\begin{equation}
t_{\text{trip,neu}} \approx n \cdot T_{s,i}(t) = \frac{n}{1 + k D_i(t)}
\end{equation}

Assuming the disturbance is primarily due to current deviation, the expression becomes:

\begin{equation}
t_{\text{trip,neu}} \propto \frac{1}{1 + k' |I - I_{\text{nom}}|}
\end{equation}

\noindent \textbf{Remark 1:} This expression highlights an inverse relationship between the neuromorphic tripping time and the fault current—mirroring the logic of traditional inverse-time relays. However, instead of relying on centralized measurements or synchronized clocks, this behavior is achieved using local, biologically inspired spiking logic.

\vspace{1em}
The classical inverse-time protection curve, known as the IDMT (Inverse Definite Minimum Time) curve and defined in the IEC 60255-151 standard, is given by:

\begin{equation}
T(I) = K \left( \left( \frac{I}{I_{\text{pickup}}} \right)^n - 1 \right)
\end{equation}

Here, $T(I)$ is the trip time, and $K$, $n$ are curve-specific constants. For example, a standard IEC curve uses $K = 0.14$, $n = 0.02$, and $I_{\text{pickup}} = 1.0$ per unit (pu). This expression is valid for $I > I_{\text{pickup}}$ to prevent singularity at the threshold.

To compare both approaches, we can plot trip time curves from the IDMT formula and the neuromorphic spiking formula for increasing fault current. In both cases, as fault current increases, trip time decreases sharply. This confirms that the neuromorphic framework exhibits IDMT-like timing behavior.

\vspace{0.5em}

\newpage
\subsection*{Supplementary Note 8 — Comparison Between Traveling Wave and Spike-Based Neuromorphic Protection}

With the rise of decentralized and inverter-dominated microgrids, there is a growing need for fast and reliable protection methods. Two promising approaches—\textbf{Traveling Wave (TW)} and \textbf{Spike-Based Neuromorphic Protection}—offer high-speed, communication-free fault detection. Though based on different principles, both aim to isolate faults in less than one cycle. This note compares their operation, accuracy, and implementation.

\subsubsection*{Working Principle and Modeling}

TW-based protection detects high-frequency electromagnetic waves generated during a fault. These travel along lines at nearly the speed of light. Their arrival and reflection times help locate the fault, modeled using the telegrapher’s equations:

\begin{equation}
\frac{\partial V(x,t)}{\partial x} = -L \frac{\partial I(x,t)}{\partial t} - R I(x,t)
\end{equation}
\begin{equation}
\frac{\partial I(x,t)}{\partial x} = -C \frac{\partial V(x,t)}{\partial t} - G V(x,t)
\end{equation}

Using wave travel speed $v$, the fault distance $m$ is estimated as:
\begin{equation}
m = \frac{v(t_4 - t_1)}{2}
\end{equation}

Spike-based neuromorphic protection models each DER as a Leaky Integrate-and-Fire (LIF) neuron. It processes local disturbances in voltage, current, and power into a scalar disturbance index:

\begin{equation}
D_i(t) = \alpha |\Delta V(t)| + \beta |\Delta I(t)| + \gamma |\Delta P(t)|
\end{equation}

This is converted into a spike interval:

\begin{equation}
T_{s,i}(t) = \frac{1}{1 + k D_i(t)}
\end{equation}

A spike is fired when the membrane potential exceeds a dynamic threshold. The first neuron to spike initiates circuit breaker operation—this is called \textbf{First-To-Spike (FTTS)} logic.\

\textit{Remark:} While TW methods depend on physical wave propagation, spike-based methods convert disturbance strength into spike timing, inspired by how neurons react to stimuli.\

Traveling wave protection responds within 1--2~ms as fault-induced wavefronts reach the terminals, but its effectiveness may reduce when faults occur at voltage zero-crossings or generate weak disturbances. Neuromorphic spike-based protection also achieves sub-cycle operation, where the tripping delay is inversely related to the disturbance index:

\begin{equation}
t_{\text{trip}} \propto \frac{1}{D_i(t)}.
\end{equation}

While traveling wave (TW) methods use polarity logic to determine fault direction and can be affected by reflections in meshed networks, spike-based approaches tend to identify the nearest faulted location naturally, as the closest DER experiences the largest disturbance and spikes first. TW schemes generally require GPS synchronization, wide-band sensors, and reliable communication links, which may add to deployment complexity. By contrast, spike-based systems process only local measurements and therefore avoid external synchronization or communication.

High-frequency noise can influence TW detection, whereas spike-based schemes employ adaptive thresholds, such as

\begin{equation}
V_{\text{th}}(t) = V_0 e^{-\eta(\Delta V + \Delta I + \dot{P})} + \lambda V_{\text{m}}(t),
\end{equation}

which help suppress small fluctuations while maintaining sensitivity to genuine faults.

\textit{Remark:} Adaptive thresholds enhance sensitivity to real faults while suppressing false trips during normal load changes.

\subsection*{Supplementary Note 9: Real‑Time Validation and Practical Deployment}

The proposed spike-based neuromorphic protection framework was implemented in real time on a 3-DER ring microgrid using an OPAL-RT simulator, as shown in supplementary Fig. 11 below. Standard, commercially available hardware was assumed: each DER has local voltage and current sensors and a digital controller capable of performing LIF updates at microsecond resolution. The disturbance index is computed locally, and spike outputs directly actuate breakers without communication.

\begin{figure*}
    \centering
    \includegraphics[scale=0.90]{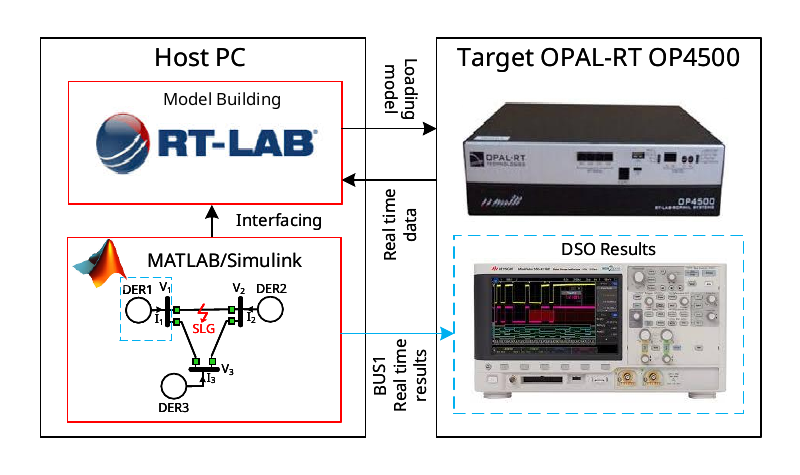}
    
    \caption*{
    \textbf{Supplementary Fig. 11:} 
    Real-time validation setup of a three-DER ring microgrid, showing MATLAB/Simulink–RT-LAB model deployment to OPAL-RT OP4500 and real-time monitoring of Bus-1 (DER-1) voltage under a Bus-1–Bus-2 fault.
}
\label{fig:control_schematic}
    \vspace{-2mm}
\end{figure*}
\begin{figure*}
    \centering
    \includegraphics[scale=0.7]{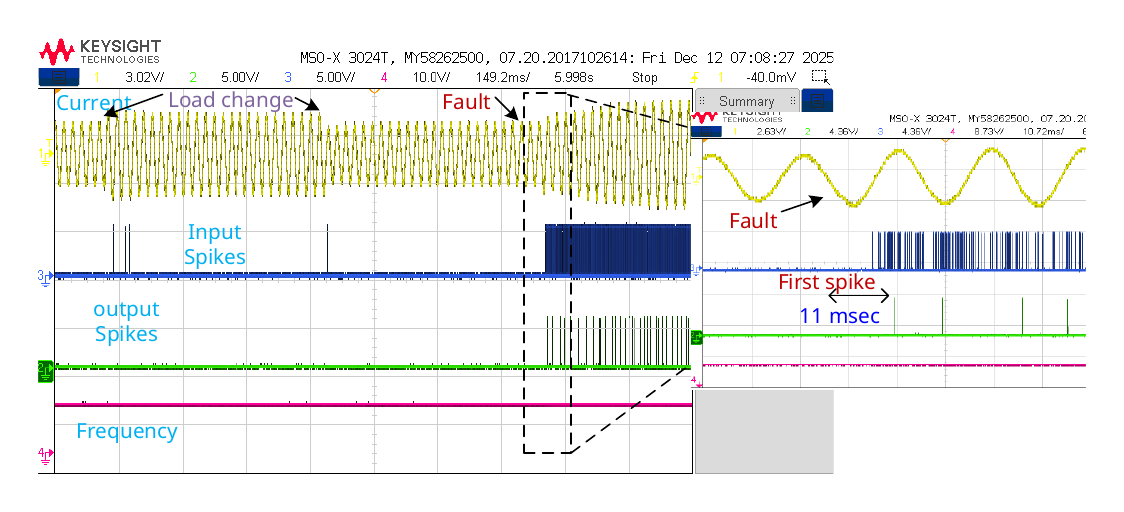}
    
    \caption*{
    \textbf{Supplementary Fig. 12: Real-time OPAL-RT validation of spike-based fault detection without circuit breaker action.} 
    Validation is performed on a three-bus, three-DER ring microgrid implemented in OPAL-RT. From top to bottom, the figure shows:  Line current at DER1 , Corresponding input spike activity, Output spike generation, System frequency.
    A load change is applied first, followed by a fault event. While the load disturbance does not trigger protection, the fault produces a sharp increase in spike activity at DER1, leading to fault detection within approximately 10 ms. The system frequency remains close to 50 Hz during detection, confirming stable operation prior to isolation.
}

    \label{fig:control_schematic}
    \vspace{-2mm}
\end{figure*}

\begin{figure*}
    \centering
    \includegraphics[scale=0.7]{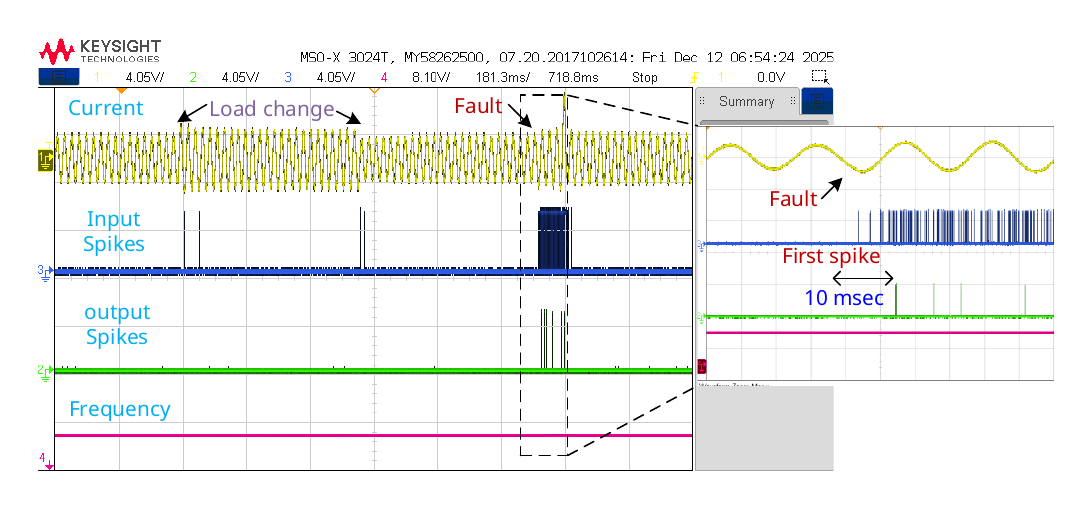}
    
   \caption*{
    \textbf{Supplementary Fig. 13: Real-time OPAL-RT validation of FTTS-based protection with circuit breaker enabled.} 
    The same three-bus, three-DER ring system is evaluated with circuit breaker logic activated. Following fault inception, DER1 emits the first spike, triggering FTTS-based breaker operation and isolating the faulted line. Line current is rapidly suppressed after tripping, and spike activity subsides accordingly. The system frequency quickly returns to nominal (50 Hz), demonstrating fast, selective isolation and effective containment of the disturbance in real time.
}

    \label{fig:control_schematic}
    \vspace{-2mm}
\end{figure*}

\newpage
Fault detection without CB actuation (Supplementary Fig. 12):
A load change is applied first, followed by a fault. The load change does not trigger spiking, whereas the fault produces a rapid spike response at the nearest DER (DER1). The fault is detected within approximately 10 ms, while the system frequency remains close to 50 Hz, confirming fast and selective fault detection without false tripping.

FTTS-based fault isolation with CB enabled (Supplementary Fig. 13):
When circuit breakers are enabled, the earliest spike from DER1 activates FTTS logic and triggers breaker operation on the faulted line. The fault is isolated rapidly, spike activity subsides after tripping, and system frequency quickly stabilizes at 50 Hz.

These real-time results closely match the simulation outcomes and confirm that the proposed neuromorphic FTTS protection achieves fast, selective fault detection and isolation under practical real-time execution.

The hardware requirements to deploy the scheme in an actual microgrid are deliberately modest and compatible with existing practice. Each DER needs only standard voltage and current measurements at its point of common coupling, sampled at tens of kilohertz, together with a digital controller (e.g., DSP, FPGA, or microcontroller integrated in today’s inverter control boards) capable of computing the disturbance index and updating the leaky‑integrate‑and‑fire neuron at microsecond‑scale time steps. The neuromorphic logic operates entirely on local measurements, so no wide‑area communication or time‑synchronization infrastructure is required; the output spike stream directly drives a local breaker or trip contact. These results confirm that the framework is practical and compatible with existing microgrid hardware.
\newpage
\subsection*{Supplementary Note 10: Theoretical Analysis and Topology-Level Validation of the Neuromorphic FTTS Protection Framework}

To provide rigorous theoretical grounding, we establish that the First-to-Spike (FTTS) mechanism always selects the DER electrically closest to the fault, independent of microgrid topology. Intuitively, the DER nearest to a fault experiences the largest disturbance index, producing the highest spike frequency and the shortest spike interval, ensuring it spikes first.

\textbf{Spike Interval Equation:}
\[
T_{s,i} = \frac{1}{1 + k D_i(t)}
\]
where $D_i(t)$ is the disturbance index at DER $i$, and $k$ is a scaling constant. Larger disturbances lead to smaller $T_{s,i}$, so the nearest DER spikes first. As the simulation step $\Delta t \to 0$, the probability of a unique first spike approaches 1, providing a topology-independent guarantee for FTTS convergence (Theorem~1, Methods~4.2).

\textbf{Spike-Latency Bounds:}
\[
T_s \in \Bigg[\frac{1}{1 + k D_{\max}}, \;\frac{1}{1 + k D_{\min}}\Bigg]
\]
where
\[
D_{\max} \propto \frac{1}{Z_{\min}}, \quad D_{\min} \propto \frac{1}{Z_{\max}}
\]

Here, $Z_{\min}$ and $Z_{\max}$ are the minimum and maximum electrical impedances from the DERs to the fault, corresponding to the closest and farthest DERs, respectively. These bounds apply to all topologies, including radial, ring, meshed, multi-ring, and dynamically islanded networks.

\textbf{Empirical Validation Across Topologies:} To confirm that FTTS behavior is topology-agnostic, the framework was evaluated on multiple microgrid structures of increasing complexity, including radial, ring, and meshed networks with different numbers of buses and DERs (see \textbf{Supplementary Figs. 8, 9, and 10}). Across all configurations, the following properties were consistently observed:

\begin{itemize}
    \item \textbf{Spatial Selectivity:} The DER closest to the fault always produced the first spike, independent of network size or interconnection pattern (Supplementary Figs. 9–10).
    \item \textbf{Scalability:} Adding buses or DERs did not alter the decision logic; FTTS selection remained purely local and event-driven (Supplementary Figs. 8–10).
    \item \textbf{Stability Under Non-Fault Events:} Load changes produced bounded disturbances that failed to trigger premature spiking due to adaptive thresholding (Supplementary Figs. 8–10).
\end{itemize}

These observations confirm that the proposed protection mechanism scales naturally with network topology and does not rely on topology-specific coordination rules.

\newpage
\subsection*{Supplementary Note 11: Fault Ride-Through (FRT) Compatibility}

The proposed neuromorphic FTTS-based protection framework is primarily developed for islanded AC microgrids, where fast and selective isolation of internal faults is essential to maintain voltage and frequency stability and to prevent inverter shutdown. In such operating modes, fault ride-through (FRT) obligations mandated by grid codes for grid-connected DERs do not strictly apply, and protection objectives prioritize rapid disconnection over continued operation during abnormal conditions.

However, to examine the robustness of the proposed scheme under grid-connected disturbances, a dedicated validation was carried out using a two-bus radial microgrid with two inverter-based DERs, connected to the main grid at Bus 1, acting as the point of common coupling (PCC) (Supplementary Fig. 14). The test scenario included a sequence of events representative of practical operation: a 30\% load change, a moderate grid-side voltage sag, and an internal single-line-to-ground (SLG) fault within the microgrid.

During the grid-side voltage sag, the quadrature-axis current naturally increases to support the load, while the neuromorphic disturbance index remains bounded. As a result, the membrane potential of the local neuron dynamics stays below the adaptive threshold, no spikes are generated, and circuit breakers remain closed. This confirms that the proposed FTTS protection does not operate spuriously during mild grid disturbances, such as voltage sags that typically fall within FRT envelopes.

In contrast, when an internal SLG fault occurs, the disturbance index rises sharply, leading to rapid spike generation and correct circuit-breaker actuation (Supplementary Fig. 15). This behavior demonstrates that the protection reliably discriminates between grid-side voltage sags and internal microgrid faults, ensuring fast isolation only when required.

These observations indicate that, while the proposed neuromorphic protection is not explicitly designed to enforce full grid-code FRT compliance, it exhibits inherent robustness to moderate voltage sags in grid-connected operation. Extending the adaptive threshold formulation to guarantee full FRT compliance under all grid-connected fault scenarios is feasible and will be addressed as part of future work.

\begin{figure*}
    \centering
    \includegraphics[scale=0.98]{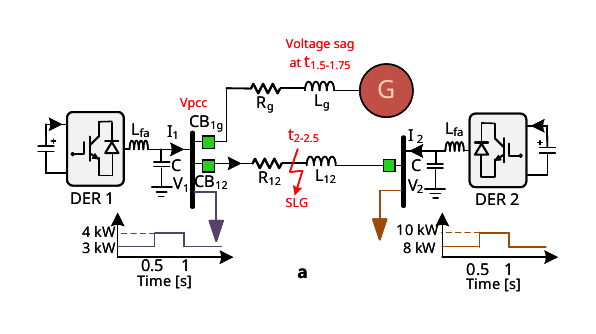}
    
   \caption*{
    \textbf{Supplementary Fig. 14: Grid-connected test system for evaluating robustness of the neuromorphic protection.} 
    Schematic of a two-bus radial microgrid with two inverter-based DERs, where Bus 1 is connected to the main grid and serves as the point of common coupling (PCC). The setup is used to assess the response of the proposed neuromorphic FTTS protection under grid-side disturbances and internal faults in grid-connected operation.
}

\newpage
    \label{fig:control_schematic}
    \vspace{-2mm}
\end{figure*}

\begin{figure*}
    \centering
    \includegraphics[scale=0.82]{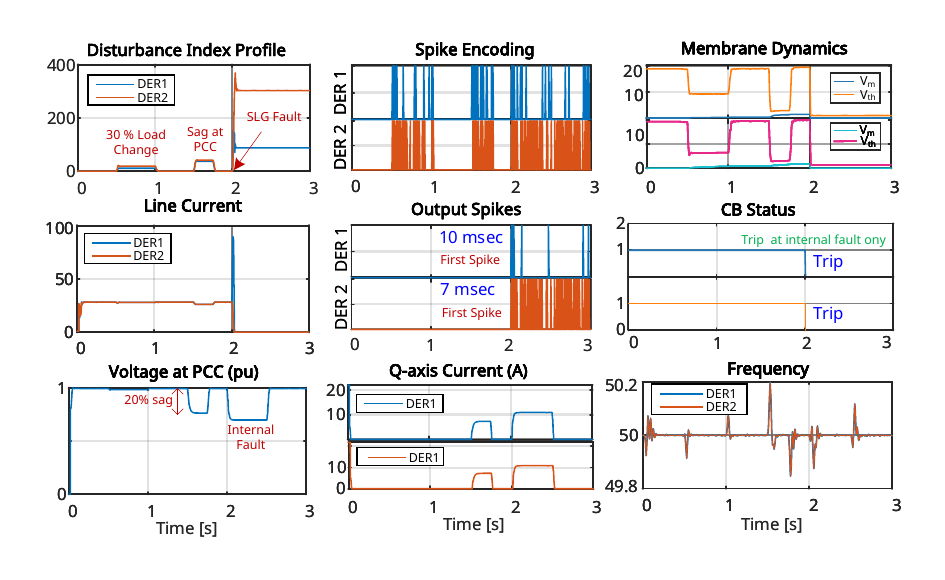}
    
    \caption*{
    \textbf{Supplementary Fig. 15: Neuromorphic protection response under grid-connected operation with load change, voltage sag, and internal fault.} 
    Time-domain results showing: A 30\% load change and a 20\% grid-side voltage sag do not trigger spiking or CB operation, while an internal SLG fault produces a sharp disturbance increase, rapid spike generation, and correct fault isolation, demonstrating selective operation and robustness of the proposed scheme.
}

    \label{fig:control_schematic}
    \vspace{-2mm}
\end{figure*}

\newpage

\subsection*{Supplementary Note 12: Performance of the proposed neuromorphic protection under heterogeneous DER ratings and control dynamics}

Practical microgrids commonly comprise distributed energy resources (DERs) with non-identical power ratings, filter parameters, and droop control gains, resulting in unequal steady-state power sharing and heterogeneous dynamic responses. To examine the robustness of the proposed neuromorphic protection framework under such realistic conditions, additional case studies are conducted using heterogeneous DER configurations in both the 3-DER and 4-DER ring microgrids.

\subsubsection*{Case 1: Heterogeneous 3-DER ring topology with CB enabled}

A 3-DER ring microgrid is considered as shown in below Fig. 16a in which each DERs have different active power droop coefficients, DC-link voltages, and output filter parameters, as summarized in Supplementary Table~5 as shown below. Specifically, DER2 is assigned a smaller droop coefficient than DER1, while DER3 is assigned a larger droop coefficient, producing unequal steady-state power sharing and different transient current responses.

\begin{table}[ht]
\caption*{\textbf{Supplementary Table 5:} Electrical and control parameters for the heterogeneous 3-DER ring topology (CB enabled).}
\label{tab:ring_microgrid_heterogeneous}
\centering
\renewcommand{\arraystretch}{1.25}
\begin{tabular}{|l|c|c|c|c|}
\hline
\textbf{Parameter (Units)} & \textbf{Symbol} & \textbf{DER1} & \textbf{DER2} & \textbf{DER3} \\
\hline
Line Voltage (V, L--L) & $V_L$ & 415 & 415 & 415 \\
Frequency (Hz) & $f$ & 50 & 50 & 50 \\
Active power droop coefficient (pu/Hz) & $m_p$ & $1.0\times10^{-4}$ & $0.9\times10^{-4}$ & $1.25\times10^{-4}$ \\
DC-link voltage (V) & $V_{\text{dc}}$ & 800 & 1000 & 1000 \\
Filter inductance (mH) & $L_f$ & 4.0 & 4.3 & 3.8 \\
Filter capacitance ($\mu$F) & $C_f$ & 200 & 220 & 180 \\
\hline
Impedance DER1--DER2 ($\Omega$) & $R_{12}+jX_{12}$ & \multicolumn{3}{c|}{$0.5 + j1.553$} \\
Impedance DER1--DER3 ($\Omega$) & $R_{13}+jX_{13}$ & \multicolumn{3}{c|}{$0.35 + j5.435$} \\
Impedance DER2--DER3 ($\Omega$) & $R_{23}+jX_{23}$ & \multicolumn{3}{c|}{$1.8 + j4.23$} \\
\hline
\end{tabular}
\end{table}

\begin{figure*}
    \centering
    \includegraphics[scale=0.72]{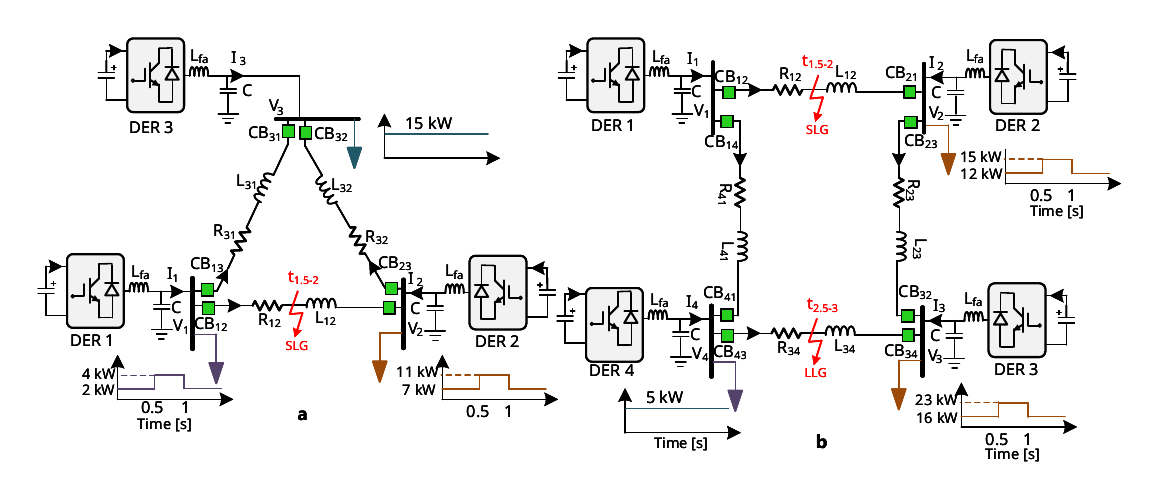}
    
    \caption*{
    \textbf{Fig. 16: Heterogeneous ring microgrid test systems.} 
\textbf{Topology:} A 3-DER ring microgrid with circuit breakers enabled (Fig.~16a) and a 4-DER ring microgrid with circuit breakers disabled (Fig.~16b). 
}

    \label{fig:control_schematic}
    \vspace{-2mm}
\end{figure*}

Simulation results, as shown in Supplementary Fig.~17 below, demonstrate each DER contributes a distinct level of active power in steady state in accordance with its droop coefficient. When a load change is applied between $0.5$ and $1~\mathrm{s}$, the disturbance index exhibits moderate and distributed variations across the DERs without activating protection mechanisms. 

In contrast, a three-phase-to-ground (LLLG) fault introduced between DER1 and DER2 from $1.5$ to $2~\mathrm{s}$ produces a sharp increase in the disturbance index at DER1 and DER2, resulting in faster spike generation and earlier membrane threshold crossing at these units. Consequently, output spikes from DER1 and DER2 initiate the tripping of CB12 and CB21, isolating the faulted line within approximately one cycle, while DER3 remains unaffected. Overall, these results confirm that spatial selectivity and rapid fault isolation are maintained despite heterogeneous droop coefficients and converter parameters.

\begin{table}[ht]
\caption*{\textbf{Supplementary Table 6:} Electrical and control parameters for the heterogeneous 4-DER ring topology (CB disabled).}
\label{tab:ring_microgrid_4der}
\centering
\renewcommand{\arraystretch}{1.25}
\begin{tabular}{|l|c|p{1cm}|p{1cm}|p{1cm}|p{1cm}|}
\hline
\textbf{Parameter (Units)} & \textbf{Symbol} & \textbf{DER1} & \textbf{DER2} & \textbf{DER3} & \textbf{DER4} \\
\hline
Line Voltage (V, L--L) & $V_L$ & 415 & 415 & 415 & 415 \\
Frequency (Hz) & $f$ & 50 & 50 & 50 & 50 \\
Active power droop coefficient (pu/Hz) & $m_p$ & $1.0\times10^{-4}$ & $0.9\times10^{-4}$ & $1.25\times10^{-4}$ & $1.5\times10^{-4}$ \\
DC-link voltage (V) & $V_{\text{dc}}$ & 1000 & 1000 & 800 & 1000 \\
Filter inductance (mH) & $L_f$ & 4.0 & 4.2 & 3.8 & 4.4 \\
Filter capacitance ($\mu$F) & $C_f$ & 200 & 220 & 180 & 210 \\
\hline
Impedance DER1--DER2 ($\Omega$) & $R_{12}+jX_{12}$ & \multicolumn{4}{c|}{$0.75 + j1.456$} \\
Impedance DER2--DER3 ($\Omega$) & $R_{23}+jX_{23}$ & \multicolumn{4}{c|}{$1.34 + j3.12$} \\
Impedance DER3--DER4 ($\Omega$) & $R_{34}+jX_{34}$ & \multicolumn{4}{c|}{$0.86 + j2.46$} \\
Impedance DER4--DER1 ($\Omega$) & $R_{41}+jX_{41}$ & \multicolumn{4}{c|}{$0.57 + j1.63$} \\
\hline
\end{tabular}
\end{table}

\begin{figure*}[ht]
    \centering
    \includegraphics[scale=0.82]{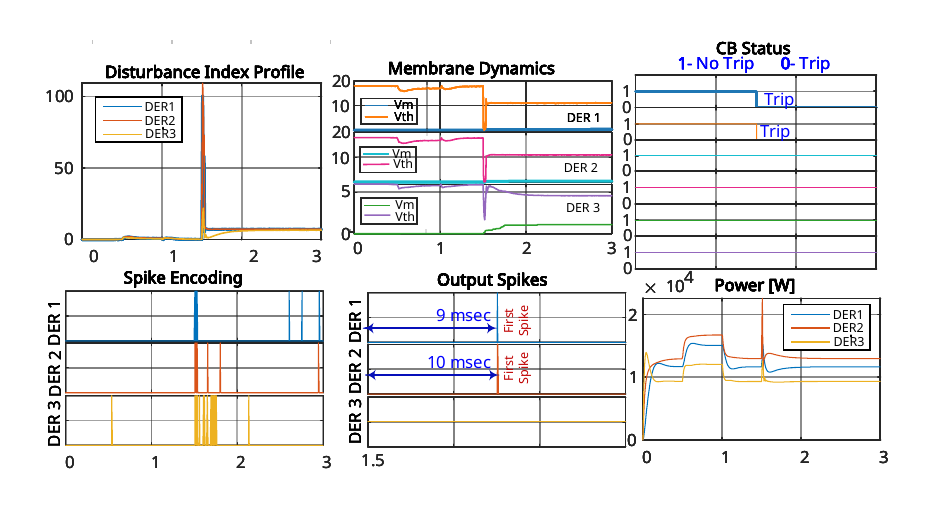}
    
    \caption*{
    \textbf{Fig. 17: Neuromorphic protection in a heterogeneous 3-DER ring microgrid with circuit breakers enabled.} 
\textbf{Disturbance sequence:} A load change occurs between 0.5--1.0 s, followed by a three-phase-to-ground (LLLG) fault between DER1 and DER2 from 1.5--2.0 s. 
\textbf{Power sharing:} DERs supply unequal steady-state active power according to their droop coefficients. 
\textbf{Selectivity:} The load change produces only mild disturbance-index variations and does not trigger protection. 
\textbf{Localization:} The LLLG fault causes a sharp disturbance-index rise at DER1 and DER2, leading to earlier spiking at these units. 
\textbf{Protection action:} Output spikes from DER1 and DER2 trigger CB12 and CB21, isolating the faulted line within approximately one cycle, while DER3 remains unaffected.
}
    \label{fig:control_schematic}
    \vspace{-2mm}
\end{figure*}

\subsubsection*{Case 2: Heterogeneous 4-DER ring topology with CB disabled}

To further evaluate scalability and diversity, a 4-DER ring microgrid as shown in Fig 1b is studied in which DERs have different droop coefficients, DC-link voltages, and filter parameters, as summarized in Supplementary Table~2 shown below. In this case, circuit breakers are disabled to examine spike-level behavior only.

Simulation results shown below in Supplementary Fig.~18 show that DERs inherently share different levels of steady-state active power in accordance with their respective droop coefficients. A single-line-to-ground (SLG) fault applied between $1.5$ and $2~\mathrm{s}$ leads to moderate spike activity that remains localized near the faulted line. When a subsequent three-phase-to-ground (LLLG) fault occurs between $2.5$ and $3~\mathrm{s}$, the higher disturbance index results in denser and faster spike generation -- with less inter-spike intervals. In both cases, the DERs electrically closest to the fault are the first to spike, despite differences in their ratings and control parameters. Even with circuit breakers disabled, frequency and voltage deviations stay localized, and the timing of the spikes continues to provide a clear indication of the fault location.

These two heterogeneous case studies demonstrate that the proposed neuromorphic protection framework does not depend on identical DER parameters or uniform control design. Instead, protection decisions are governed by the relative timing of locally generated spikes, which are driven by measurable physical disturbances. Heterogeneity in droop coefficients, DC-link voltages, and filter parameters therefore does not degrade performance but rather reinforces the suitability of the proposed approach for heterogeneous inverter-dominated microgrids.

\newpage
\begin{figure*}
    \centering
    \includegraphics[scale=0.82]{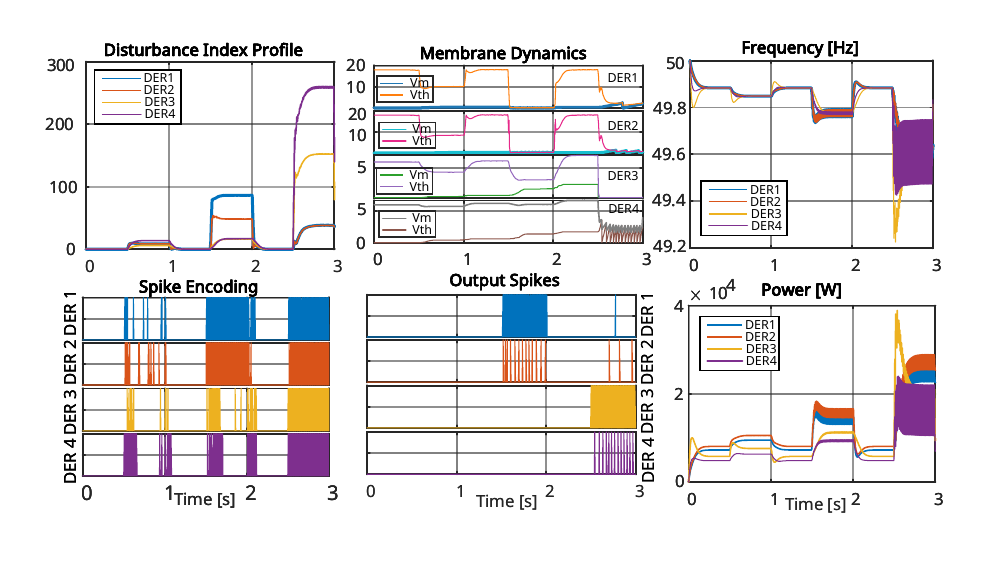}
    
    \caption*{
    \textbf{Fig. 18: Spike-based fault localization in a heterogeneous 4-DER ring microgrid with circuit breakers disabled.} 
\textbf{Disturbance sequence:} A single-line-to-ground (SLG) fault is applied from 1.5--2.0 s, followed by a three-phase-to-ground (LLLG) fault from 2.5--3.0 s. 
\textbf{Power sharing:} DERs inherently share different levels of steady-state active power in accordance with their droop coefficients. 
\textbf{Severity awareness:} The LLLG fault produces denser and faster spiking than the SLG fault. 
\textbf{Localization:} DERs electrically closest to the fault are the first to spike despite heterogeneous parameters. 
\textbf{Resilience without tripping:} Even with circuit breakers disabled, voltage and frequency deviations remain localized, and spike timing clearly indicates the faulted line.

}

    \label{fig:control_schematic}
    \vspace{-2mm}
\end{figure*}

\newpage

\newpage
\subsection*{Supplementary Note 13: Impact of DER Overcurrent Limiting on Neuromorphic Fault Detection}

Grid-forming DERs typically employ fast current-limiting mechanisms to prevent semiconductor overcurrent during severe disturbances. Under such operation, the inverter output current is rapidly clamped to a maximum allowable value \( I_{\text{max}} \), while the terminal voltage experiences a pronounced sag and the delivered active power collapses. In the proposed framework, the disturbance index is constructed as a weighted aggregation of voltage, current, and active-power deviations, such that

\[
D(t) = \alpha \left| \Delta V(t) \right| + \beta \left| \Delta I(t) \right| + \gamma \left| \Delta P(t) \right|.
\]

When current limiting is active, \( \left| \Delta I(t) \right| \) becomes bounded, but both \( \left| \Delta V(t) \right| \) and \( \left| \Delta P(t) \right| \) increase sharply due to voltage collapse and power imbalance. Consequently, the overall disturbance index still rises substantially, driving rapid membrane potential growth and compressed spike intervals. This ensures that fault-induced spiking remains strong even when the current channel saturates.

To explicitly address this issue, a standard current-limiting block was implemented in the grid-forming control of all DERs in both heterogeneous test systems shown in Fig.~1, namely the 3-DER ring microgrid with circuit breakers enabled (Fig.~16a) and the 4-DER ring microgrid with circuit breakers enabled (Fig.~16b). The limiter constrains the inverter output current to a maximum value $I_{\max}$ by saturating the reference current once $|I^*| > I_{\max}$. This structure closely emulates the protection strategy adopted in practical grid-forming inverters.

\subsubsection*{Case 1: 3-DER Ring Microgrid with Current Limiting Strategy}

For the 3-DER ring microgrid, simulation results with current limiting activated are presented in Fig.~19. A load change is applied between $0.5$--$1.0~\mathrm{s}$, followed by an SLG fault between DER1 and DER2 at $1.5~\mathrm{s}$. After fault inception, the line current rises briefly and is rapidly clamped at $I_{\max}$. Nevertheless, a pronounced voltage sag appears at buses adjacent to the fault, and the active power outputs of DER1 and DER2 drop sharply due to the voltage collapse and network imbalance. These combined effects produce a strong rise in the disturbance index at DER1 and DER2, which in turn leads to accelerated spike encoding, rapid membrane potential growth, and threshold crossing approximately $13~\mathrm{ms}$ after fault inception. The resulting output spikes initiate the tripping of CB12 and CB21, isolating the faulted line, while DER3 exhibits only mild disturbance and remains non-tripping. This demonstrates that even under strictly limited fault current, voltage and power deviations alone are sufficient to trigger reliable neuromorphic detection.

\begin{figure*}
    \centering
    \includegraphics[scale=0.90]{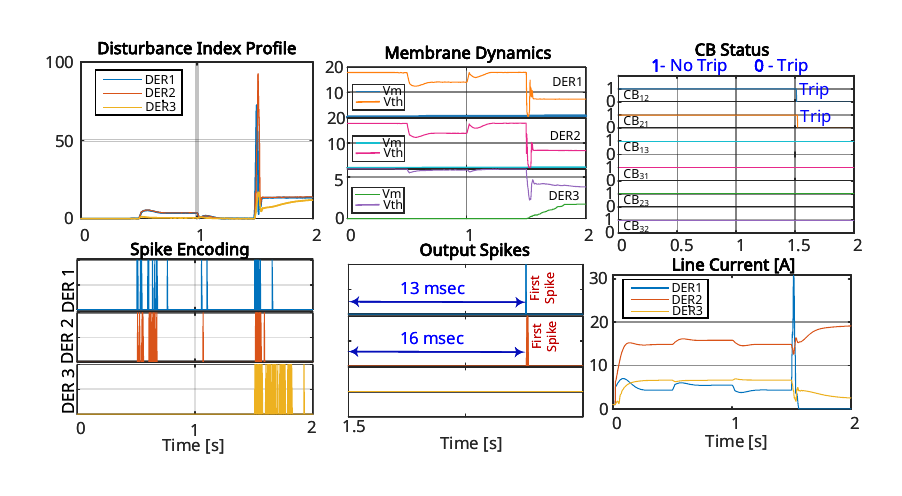}
    
    \caption*{
    \textbf{Fig. 19: Fault detection under inverter current limiting in a heterogeneous 3-DER ring microgrid.}
    \textbf{Disturbance sequence:} A load change is applied from $0.5$--$1.0~\mathrm{s}$, followed by an SLG fault between DER1 and DER2 at $1.5~\mathrm{s}$ with inverter output currents constrained by a current limiter.
    \textbf{Robustness to current saturation:} Although line currents are rapidly clamped to their maximum limits after fault inception, pronounced voltage sag and active-power imbalance generate a sharp rise in the disturbance index at DER1 and DER2.
    \textbf{Neuromorphic response:} The elevated disturbance index produces accelerated spike encoding, faster membrane potential growth, and first spikes at approximately $13~\mathrm{ms}$ after fault inception.
    \textbf{Selective isolation:} Output spikes from DER1 and DER2 trigger CB12 and CB21, isolating the faulted line, while DER3 remains unaffected—demonstrating reliable protection despite current-limited operation.
    }
    
    \label{fig:fig4_current_limit}
    \vspace{-2mm}
\end{figure*}

\subsubsection*{Case 2: 4-DER Ring Microgrid with Current Limiting Strategy}

\begin{figure*}
    \centering
    \includegraphics[scale=0.82]{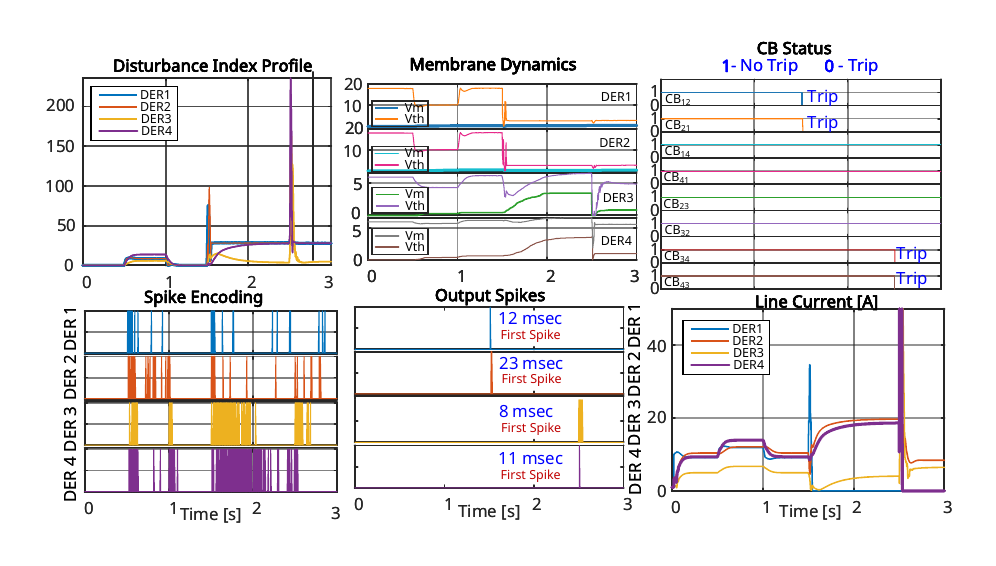}
    
    \caption*{
    \textbf{Fig. 20: Multi-event fault response with inverter current limiting in a heterogeneous 4-DER ring microgrid.}
    \textbf{Disturbance sequence:} A load change occurs between $0.5$--$1.0~\mathrm{s}$, an SLG fault is introduced between DER1 and DER2 from $1.5$--$2.0~\mathrm{s}$, and a more severe LLLG fault is applied between DER3 and DER4 from $2.5$--$3.0~\mathrm{s}$, with all DERs operating under current-limited conditions.
    \textbf{Severity awareness:} Despite current saturation, the LLLG fault produces faster and denser spiking than the SLG fault, reflecting higher disturbance intensity.
    \textbf{Spatial selectivity:} DERs electrically closest to each fault location are the first to spike, independent of heterogeneous ratings and control parameters.
    \textbf{Practical relevance:} Correct and timely spiking under current-limited operation confirms that the neuromorphic framework remains effective for realistic grid-forming inverters.
    }
    
    \label{fig:fig5_multi_event}
    \vspace{-2mm}
\end{figure*}

For the 4-DER ring microgrid with current limiting, results are shown in Fig.~20, where a load change occurs between $0.5$--$1.0~\mathrm{s}$, an SLG fault between DER1 and DER2 from $1.5$--$2.0~\mathrm{s}$, and a subsequent LLLG fault between DER3 and DER4 from $2.5$--$3.0~\mathrm{s}$. In both fault events, inverter currents are rapidly saturated at $I_{\max}$. Despite this, the SLG fault produces a moderate voltage sag and power imbalance that yields first spikes at approximately $12~\mathrm{ms}$, whereas the more severe LLLG fault causes deeper voltage depression and larger active-power deviation, resulting in denser spiking and faster first-spike generation at approximately $8~\mathrm{ms}$. These results confirm that fault severity continues to be reflected in spike timing and density, even when current magnitude is constrained.

From an analytical standpoint, under current-limited operation the incremental current deviation is bounded, whereas voltage deviation increases significantly during faults and the active-power deviation
\[
\Delta P(t) = V(t) I(t) - V_0 I_0
\]
(where \( V_0 \) and \( I_0 \) denote the corresponding steady-state pre-disturbance voltage and current operating points of the DER) becomes large because \( V(t) \) collapses while \( I(t) \) remains clamped. As a result, the disturbance index remains dominated by the voltage and power terms, ensuring a substantial increase during faults even if the current term is saturated. This multi-channel construction is the fundamental reason for the robustness of the proposed method to current limiting.

Overall, these additional studies demonstrate that inverter current limiting does not degrade the proposed neuromorphic fault detection framework. Instead, the reliance on voltage and power transients makes the method inherently compatible with grid-forming DERs employing protective current saturation. This further strengthens the practical relevance and applicability of the proposed approach in inverter-dominated microgrids.

\newpage

\nocite{kasi2021energy,shaban2021adaptive,salaj2020spike,goltz2021fast,martin2021eqspike,sahoo2024spike,yang2022lead,10445758,diao2024inferring,ganguly2024spike,baronig2024advancing,rishu2024simulation,jhingeffect,reda2022effect,reda2023multi,schneider2019overcurrent,etingov2022ai,wilches2021survey,liu2024evaluation,paruthiyil2025travelling,jimenez2023protection,IEEE_C37_112_1996,IEEE_1547_2018,ENTSOE_NC_RfG_2016,CEA_CERC_GridCode_2023}
\newpage
\printbibliography[title={Supplementary References}]
\end{refsection}


\end{document}